\documentclass[aoas]{imsart}

\RequirePackage{amsfonts,amsmath,amssymb,amsthm}
\RequirePackage[authoryear]{natbib}
\RequirePackage[colorlinks,citecolor=blue,urlcolor=blue]{hyperref}
\usepackage[pdftex]{graphicx}
\DeclareGraphicsExtensions{.png,.pdf,.jpg}
\usepackage{bm}
\usepackage{multirow}
\usepackage{longtable}
\usepackage{mathtools}

\startlocaldefs
\theoremstyle{plain}

\theoremstyle{remark}

\endlocaldefs

\begin{document}

\begin{frontmatter}
\title{Bayesian Segmentation Modeling of Epidemic Growth}
\runtitle{Bayesian Segmentation Modeling of Epidemic Growth}

\begin{aug}
\author[A]{\fnms{Tejasv}~\snm{Bedi}\ead[label=e1]{tejasv.bedi@utdallas.edu}},
\author[B]{\fnms{Yanxun}~\snm{Xu}\ead[label=e2]{yanxun.xu@jhu.edu}}
\and
\author[A]{\fnms{Qiwei}~\snm{Li}$^*$\ead[label=e3]{qiwei.li@utdallas.edu}}
\address[A]{Department of Mathematical Sciences, The University of Texas at Dallas\printead[presep={,\ }]{e1,e3}}

\address[B]{Department of Applied Mathematics and Statistics, Johns Hopkins University\printead[presep={,\ }]{e2}}
\end{aug}

\begin{abstract}
Tracking the spread of infectious disease during pandemics has posed an enormous challenge to both the public and private sectors on a global scale. To facilitate informed public health decision-making, concerned parties usually rely on short-term daily and weekly projections generated via predictive modeling. Several deterministic and stochastic epidemiological models, including growth and compartmental models, have been proposed in the literature. Such models assume that an epidemic would last over a short duration and the observed cases/deaths would attain a single peak; however, some infectious diseases, such as COVID-19, extend over a longer duration than expected. Moreover, government interventions have led to time-varying disease transmission rates, making the observed data multi-modal. To address these challenges, we propose stochastic epidemiological models under a unified Bayesian framework augmented by a change-point detection mechanism to account for multiple peaks. The Bayesian framework allows us to incorporate prior knowledge, such as dates of influential policy changes, to precisely predict change-point locations. We develop a trans-dimensional reversible jump Markov chain Monte Carlo algorithm to sample the posterior distributions of epidemiological parameters while estimating the number of change points, as well as the resulting parameters. We evaluate our proposed method and compare it to alternative methods in terms of change-point detection, parameter estimation, and long-term forecasting accuracy on both simulated and COVID-19 data from major states in the United States.
\end{abstract}

\begin{keyword}
\kwd{change-point detection}
\kwd{COVID-19}
\kwd{final epidemic size}
\kwd{reversible jump Markov chain Monte Carlo}
\kwd{stochastic growth model}
\end{keyword}

\end{frontmatter}

\section{Introduction}
\label{sec:intro}

 


As improving healthcare systems result in decreasing mortality rates, there is growing epidemiological interest in recognizing the impact of diseases that place a burden on human health \citep{vos2016global}. Infectious diseases, including those caused by agents such as the human immunodeficiency virus (HIV) (1981), severe acute respiratory syndrome coronavirus 1 (SARS-CoV-1) (2002), influenza A virus subtype H1N1 (2009), Middle East respiratory syndrome-related coronavirus (MERS-CoV) (2012), Ebola virus (2013), and severe acute respiratory syndrome coronavirus-2 (SARS-CoV-2) (2019) have directly impacted human health outcomes over the past four decades \citep{roychoudhury2020viral}. Statistical modeling, simulations, and inferences are key to assessing the impact of epidemics on populations and guiding public policy and healthcare systems \citep{chowell2017fitting}.


We focus on the impact of SARS-CoV-2, which has caused widespread infections, hospitalizations, and deaths by way of a global pandemic. COVID-19, the disease caused by SARS-CoV-2, was first reported in Wuhan, China in December 2019 \citep{doi:10.1056/NEJMoa2001316}. In the initial phase of the pandemic from January to May 2020, governments practiced non-pharmaceutical interventions (NPIs) to reduce the spread of the virus, including lockdowns, social distancing, school closures, and public event bans. After several attempts at modeling the impact of NPIs, researchers noted that major interventions, including lockdowns and large gathering bans, were most effective at reducing disease spread in the pandemic's early phase \citep{flaxman2020estimating, soltesz2020effect, banholzer2021estimating}. Since the end of the first wave of infections, rapid pharmaceutical advancements have been deployed in most countries, chiefly in the form of vaccinations that have proven safe and efficacious in preventing most symptomatic and asymptomatic infections \citep{covid2021covid}. However, the emergence of new viral variants (e.g., Delta and Omicron) has led to additional waves in countries including Brazil, India, and the United States (U.S.) among both unvaccinated and vaccinated individuals, sparking concerns about waning vaccine efficacy \citep{nowroozi2021severe, smith2021transmission, alexandar2021comprehensive, zhang2021transmission, lopez2021effectiveness, thangaraj2021predominance}. Moreover, this surge of new mutations has made the development of future vaccines and testing procedures challenging \citep{wang2020mutations}. In addition, other difficulties have frustrated the modeling and forecasting of disease data, which is ever-changing and influenced by many external factors, including time-varying transmission rates, mobility, interventions, and mutations. The resultant data is multi-modal, with varying growth trajectories within specific time windows. Hence, we define four goals that logically follow from said difficulties: 1) Finding optimal criteria for dividing the entire data into heterogeneous sub-intervals within which the disease dynamics are stable; 2) Estimating the location and number of change-points (i.e., time points that cause a structural shift in growth trajectories of the underlying data); 3) Estimating the total number of cases, deaths and hospitalizations by the end of the epidemic, for a particular region, given the dynamic nature of the data; and 4) Performing long-term forecasting given the shifting dynamics of the data. This paper aims to simultaneously address the aforementioned goals with an approach based solely on time series observations of cumulative infections, which are the most accessible form of disease data during a pandemic.


We define a change-point as a sudden variation of significant importance in time series data. It is usually attributed to structural changes caused by certain significant events. Change-point estimation revolves around inferring the causes of changes that were recorded \citep{aminikhanghahi2017survey}; therefore, we are not only interested in the locations of change-points, but also the causes associated with them. There are many applications of change-point analysis, including speech processing, climatology, and image analysis \citep{truong2020selective}. 
In the context of epidemiology, \cite{yu2013change} proposed a change-point detection procedure for the susceptible-infected-removed (SIR) model via binomial thinning processes as applied to 2001-2012 influenza data. Other approaches incorporate thresholding parameters within generalized linear regression frameworks for evaluating vaccine-associated changes on hospitalizations related to pneumococcal infections \citep{brilleman2017bayesian}, and analyzing delays in disease reporting of acquired immunodeficiency syndrome (AIDS) \citep{brookmeyer1990analysis, tabnak2000change}. 

Since the emergence of COVID-19, change-point analysis is gaining increasing attention in the context of disease modeling and epidemiology. Government interventions (e.g., rapid COVID-19 testing, lockdowns, mask mandates, and vaccinations) have led to dynamic disease transmission rates that result in drastically varying growth trajectories of COVID-19 cases, hospitalizations, and deaths across different regions. For instance, \cite{dehning2020inferring} and  \cite{Jiang2020.10.06.20208132} attempted to jointly model change-points and key epidemiological features (including time-varying growth rates and basic reproduction numbers) via compartmental models. On the other hand, other works have implemented piece-wise linear regression models and data-driven approaches for change-point detection in COVID-19 data \citep{jiang2020time,kuchenhoff2021analysis, pedersen2021data}. Piece-wise linear regression models have been extended to longitudinal studies in other contexts, such as piece-wise growth models (PGM) \citep{chou2004piecewise, kohli2013modeling} and piece-wise growth mixture models (PGMM) \citep{kohli2015fitting, lock2018detecting} that stratify data based on linear growth or decline. The location or number of change-points in those models require pre-specification. Moreover, each change-point indicates a shift in linear growth rates between two segments, resulting in a non-linear curve over the entire time series.

To overcome these limitations and track the spread of infectious disease based on minimal data, we propose a piece-wise generalized growth model that we call Bayesian segmentation modeling of epidemic growth (BayesSMEG). BayesSMEG estimates not only the number of change-points and their locations, but also key parameters of epidemiological interest, such as final epidemic sizes. Apart from accounting for non-linearity across the entire time series, BayesSMEG allows sub-exponential growth or decay within each segment, resulting in a smooth curve fit without necessarily increasing the complexity of the model by excessive stratification or segmentation. This enables the definition of change-points based on shifts within epidemic waves. Furthermore, our framework directly models case numbers via a negative binomial (NB) model, accounting for measurement errors and any unmeasurable randomness \citep{li2021evaluating}. To the best of our knowledge, BayesSMEG is unique among statistical methods in its ability to estimate the number and locations of change-points whilst characterizing transmission dynamics through a growth model.

This paper is structured as follows. In Section \ref{sec:back}, we provide an overview of growth curves that form the foundation of the proposed model. In Sections \ref{sec:manual} and \ref{sec:automatic}, we discuss the modeling framework and two corresponding model-fitting strategies when the number of change-points is respectively known or unknown. To generate point and interval estimates, the posterior inference process is detailed in Section \ref{sec:post}. We further evaluate the accuracy of change-point detection and epidemiological parameter estimation via simulation studies in Section \ref{sec:sim}. In Section \ref{sec:res} we perform change-point analysis and long-term forecasting of daily cases in California and New York. We conclude the paper and discuss future work in Section \ref{sec:conc}.

\section{Background Material}
\label{sec:back}

In this section, we present an overview of growth curves, which will be later augmented by a change-point indicator vector, resulting in a piece-wise growth model. 
Supplementary Table S1 lists the key notation used throughout this paper.

\subsection{Data notation}

	Let $\bm{C}=(C_1,\cdots,C_T)^\top$ be a sequence of cumulative confirmed case numbers observed at $T$ successive equally-spaced time points (e.g., days or weeks) in a closed region, where each entry $C_t\in\{C_0,\cdots,N\}$ for $t=1,\cdots,T$ and $C_t\ge C_{t-1}$. Here $C_0\in\mathbb{N}$ and $N\in\mathbb{N}$ denote the initial confirmed case number and population size in the region, respectively. Let $\dot{\bm{C}}=(\dot{C}_1,\cdots,\dot{C}_T)^\top$ be the lag-one difference of $\bm{C}$, where $\dot{C}_1=C_1-C_0$ and each following entry $\dot{C}_t=C_t-C_{t-1},t=2,\cdots,T$. Note that $\dot{\bm{C}}$ can be interpreted as the sequence of newly added confirmed case numbers. When analyzing a series of infectious disease data, the time series data $\bm{C}$ could also be cumulative death numbers, cumulative recovery case numbers, or their sums. In actuality, only confirmed cases and deaths are reported in most regions; recovery data are either unavailable or underreported. Therefore, this paper focuses on tracking infectious disease based solely on cumulative confirmed cases ${\bm{C}}$, or equivalently, the derived newly added confirmed cases $\dot{\bm{C}}$.

\subsection{Growth curves}	

	Growth curve fitting is a widely used modeling technique in the study of epidemic growth because it requires only a sequence of cumulative confirmed case numbers $\bm{C}$ \citep{chowell2014west,pell2018using,batista2020estimation, jia2020prediction,wu2020generalized,li2021evaluating}. An epidemic usually consists of three stages: 1) An initial exponential growth phase, 2) A middle phase having sub-exponential growth rates, and 3) A final phase characterized by diminishing rates until the epidemic finally subsides. Such a phenomenological structure can be captured by an `S'-shaped trend pattern (e.g., a logistic growth curve). In mathematical literature, it is described by a function of continuous time $u$:
 \begin{equation}
 y(u) \quad=\quad \frac{K}{1 + \exp(-\lambda(u - u_0))},\quad u \in \mathbb{R}
 \end{equation}
 where $K \in (0, N]$ denotes the final epidemic size, $\lambda\in\mathbb{R}^+$ reflects the early growth rate of the epidemic, and $u_0\in\mathbb{R}$ is the time point at which the first half of the epidemic duration has elapsed. In fact, this above growth equation is a solution of the first-order non-linear ordinary differential equation (ODE) below:
 
\begin{equation}
    \frac{dy(u)}{du} \quad = \quad \lambda y(u) \left(1 - \frac{y(u)}{K}\right).
\end{equation}
Recently, several generalized variants of the logistic growth curves have been used to model epidemiological data \citep{hsieh2004sars,hsieh2009richards,hsieh2009intervention,hsieh2009turning,hsieh2010pandemic, viboud2016generalized,chowell2017fitting, chowell2019novel,chowell2020real,covid2020forecasting,roosa2020real,li2021evaluating}. The generalized logistic curve (GLC) is an extension of the logistic curve with a growth scaling parameter, for which the ODE is:
\begin{equation}\label{ODE}
    \frac{dy(u)}{du} \quad = \quad \lambda y(u)^p \left(1 - \frac{y(u)}{K}\right).
\end{equation}
The additional parameter $p \in [0, 1]$ scales the initial growth rate, being strictly exponential for $p = 1$, sub-exponential for $0 < p < 1$, and strictly linear for $p = 0$. Although our proposed model builds upon the GLC, it can be easily extended to other types of growth curves studied in \cite{li2021evaluating}. 

The basic assumption of the underlying growth curve is that the data is uni-modal and only a single peak is attainable; however, the validity of this assumption becomes questionable if the epidemic lasts longer than expected and exhibits multiple peaks, which has been shown with COVID-19. To address these issues, we incorporate a change-point detection mechanism that detects multiple surges and drops in the data. In particular, we divide the data into $M$ segments, or sub-epidemics (using these two terms interchangeably), and characterize epidemiological dynamics via coupled growth ODEs. The following section presents a Bayesian piece-wise GLC model and the associated methodology for sampling from the posterior distribution. In Section \ref{sec:automatic}, we extend the model to the case where the number of change-points is unknown.

\section{Manual Change-point Detection Framework}\label{sec:manual} 


A change-point is a structural variation in an observable sequence that holds significant importance and is not merely a random deviation. In the context of epidemiological data, it manifests as a spike in cases indicating the possibility of another wave.
For instance, consider a change-point indicator vector $\bm{\delta}=(\delta_1,\cdots,\delta_T)^\top$ such that for all $t= 2,\cdots,T$, with $\delta_t=1$ if a change-point occurs at time $t$ and $\delta_t=0$ otherwise. We set $\delta_1 = 1$ by default, interpreting the first time point as the zeroth change-point. The change-points divide the data into $M=\sum_{t=1}^T\delta_t$ segments, such that each time point having $\delta_j=0$ is bounded between two adjacent change-points. Here, $M$ is fixed \textit{a priori}; the case when $M$ is unknown will be fully discussed in Section \ref{sec:automatic}.
We can further re-parameterize $\bm{\delta}$ into a change-point segmentation vector $\bm{z}$ such that $z_t = \sum_{t' = 1}^t \delta_{t'}$, while $\bm{\delta}$ is the lag-one difference of $\bm{z}$ (i.e., each $\delta_t = z_t - z_{t-1}$ for all $t=2,\cdots,T$). To express the time points that lie under the $m$-th segment where $m\in\{1,\cdots,M\}$, we refer to the set $\{t: z_t = m\}$.

\subsection{Data likelihood}
For modeling the newly added case number at time $t$, i.e., $\dot{C}_t$, in the $m$-th segment, we adopt a segmentation framework given by:
\begin{equation}
    \dot{C}_t|z_t=m,\bm{\Theta}_m,\phi\quad\sim\quad \text{NB}(g(C_{t-1},\bm{\Theta}_m), \phi), \quad t=1,\cdots,T,
\end{equation}
where $g(\cdot)$ is a negative binomial (NB) mean function with segment-specific epidemiological parameters denoted by $\bm{\Theta}_m$. We assume a homogeneous dispersion $\phi$ across all segments for ease of estimation. However, the framework could be extended by assuming segment-specific dispersion parameters. We further assume $\dot{\bm{C}}$ to be a stochastic process following the Markov property (i.e., the state at time $t$ only depends on the state at time $t-1$). As a result, $g(\cdot)$ is only required to be a function of $C_{t-1}$ and the segment-specific parameters to generate the new case number at time $t$. Thus, the complete data likelihood can be written as:
\begin{equation}\label{lklh}
f(\bm{\dot{C}}|\bm{\delta},\bm{\Theta}_1,\cdots,\bm{\Theta}_M,\phi) \quad=\quad f(\bm{\dot{C}}|\bm{z},\bm{\Theta}_1,\cdots,\bm{\Theta}_M,\phi)\quad=\quad \prod_{m=1}^M \prod_{\{t:z_t=m\}} f(\dot{C}_t|z_t=m,\bm{\Theta}_m,\phi),
\end{equation}
where the partial data likelihood for each observed count is an NB probability mass function:
\begin{equation}
f(\dot{C}_t|z_t=m,\bm{\Theta}_m,\phi) \quad=\quad \frac{\Gamma(\dot{C}_t+\phi)}{\dot{C}_t!\Gamma(\phi)}\left(\frac{\phi}{g(C_{t-1},\bm{\Theta}_m) + \phi}\right)^{\phi}\left(\frac{g(C_{t-1},\bm{\Theta}_m)}{g(C_{t-1},\bm{\Theta}_m)+\phi}\right)^{\dot{C}_t}.
\end{equation}
Hence, the entire duration of the epidemic is segmented into $M$ consecutive and mutually exclusive sub-epidemics, such that growth curves can be fit to each segment separately. We further formulate the NB mean function $g(\cdot)$ based on the discrete version of (\ref{ODE}):
\begin{equation}
    g(C_{t - 1}, \bm{\Theta}_m) \quad = \quad \lambda_m C_{t - 1}^{p_m} \left(1 - \frac{C_{t - 1}}{K_m}\right)
\end{equation}
for all segments $m = 1, \cdots, M$, where $\bm{\Theta}_m = \{K_m, \lambda_m, p_m\}$.

\subsection{Priors}
\label{sec:prior}
As the relationship among the parameters is unknown, prior independence is a reasonable assumption between segments and within each segment; therefore, we decompose the joint prior as $\pi(\bm{\delta}, \bm{\Theta}_1,\cdots,\bm{\Theta}_M, \phi) =\pi(\bm{\delta})\pi(\phi)\prod_{m=1}^M\pi(K_m)\pi(\lambda_m)\pi(p_m)$. 

The change-point indicator vector $\bm{\delta}$ segments the entire epidemic duration into $M$ sub-epidemics. For the sake of simplicity and to ensure that there are enough data points to model each sub-epidemic, we adopt the following assumptions: 1) $\delta_1 = 1$ by default; 2) $\delta_t = 0$ is fixed for all $t = 2,\cdots, Q$ and $t = T - (Q-1)+1,\cdots, T$ to ensure that the first and the last sub-epidemics have at least $Q$ time points; 3) For any given consecutive change-points $\delta_t = \delta_{t'} = 1$ we must have $|t - t'| \geq Q$ for all $t, t' = Q+1, \cdots, T - (Q-1)$ such that no bounded sub-epidemic has less than $Q$ time points. For the analysis of daily report data, we suggest setting $Q=7$, which corresponds to an epidemiological week. In general, we assume that each $\delta_t$ has a Bernoulli prior, $ \delta_t|\omega_t \sim \text{Bern}(\omega_t)$. Due to the above restrictions on $\bm{\delta}$, the hyper-parameters are fixed at $\omega_1 = 1$ and $\omega_t = 0$ for $t = 2,\cdots, Q$ and $t = T - (Q-1)+1,\cdots, T$, while $\omega_t=\omega_0$ (e.g., a constant small value such as $\omega_0 = 0.001$) for $t = Q+1, \cdots, T - (Q-1)$. For a specific time point $t$, if prior information is available in support of it being a change-point, then we could set $\omega_t$ a relatively large value, e.g., $\omega_t\geq 0.5$.

To complete the model specification, we choose non-informative prior settings for all segment-specific epidemiological parameters. We assume that $K_m$ follows a discrete uniform distribution within its range $[a_K, b_K]$, where $a_K=\underset{\{t:z_t = m\}}{\max}C_t$, $b_K = \lceil \rho N \rceil$, and $\rho \in (0, 1]$ is a proportionality constant that controls the proportion of total population likely to get infected in the entire duration of the epidemic. Setting $\rho = 1$ is equivalent to assuming the entire population of a particular region to be susceptible. Since $\lambda_m, \phi \in \mathbb{R}^+$, we assume that $\lambda_m \sim \text{Ga}(a_\lambda, b_\lambda)$ and $\phi \sim \text{Ga}(a_\phi, b_\phi)$, and recommend $a_\lambda = b_\lambda = a_\phi = b_\phi = 0.001$ for a weakly informative setting. As $p_m \in [0, 1]$, we assume that $p_m \sim \text{Beta}(a_p, b_p)$. Further setting $a_p = b_p = 1$ is equivalent to having a uniform prior distribution within its range.


\subsection{Posterior sampling via MCMC}
To estimate change-point locations, the posterior distribution of $\bm{\delta}$ given by $\pi(\bm{\delta}|\dot{\bm{C}},\bm{\Theta}_1,\cdots,\bm{\Theta}_M,\phi)\propto f(\dot{\bm{C}}|\bm{\delta}, \bm{\Theta}_1,\cdots,\bm{\Theta}_M, \phi) \pi(\bm{\delta})$ is of direct interest. Posterior sampling of those segment-specific epidemiological parameters also informs the growth dynamics within each sub-epidemic. For a particular segment-specific parameter, i.e., $\theta_m \in \bm{\Theta}_m=\{K_m, \lambda_m, p_m\}$, we require only the partial data likelihood in estimating its posterior density, $\pi\left(\theta_m \Big|\dot{\bm{C}},\bm{\Theta}_{m} \backslash \theta_m,\phi\right)  \propto  \prod_{\{t:z_t=m\}}f\left(\dot{{C}}_m\Big|z_t=m,\bm{\Theta}_{m}, \phi\right) \pi(\theta_m)$.
Although not of direct interest, the posterior density of $\phi$ can be obtained in a similar fashion as $\bm{\delta}$, due to the assumption of homogeneous dispersion across all segments. None of the posterior densities have a closed-form expression and are not analytically tractable; consequently, neither posterior can be sampled directly, nor are their summary statistics obtainable. Hence, posterior sampling will be based on Markov chain Monte Carlo (MCMC) approximation; in particular, a random-walk Metropolis-Hastings (RWMH) algorithm will be implemented. The MCMC algorithms are detailed in Section S1 of the supplement.

\section{Automatic Change-point Detection Framework}
\label{sec:automatic}

In this section, we introduce an additional layer of complexity to BayesSMEG's modeling framework by treating the number of sub-epidemics $M$ as an unknown parameter. This results in a trans-dimensional problem as changes in $M$ would lead to changes in the dimensionality of $\bm{\mathcal{T}}=\{\bm{\Theta}_1,\cdots,\bm{\Theta}_M\}$, where each $\bm{\Theta}_m=\{{K}_m,{\lambda}_m,{p}_m\}$. Note that the dimensionality of $\bm{\mathcal{T}}$ is $3M$. MCMC algorithms, including RWMH, do not allow trans-dimensional moves that modify the dimensionality of the parameter space; therefore, we employ reversible jump MCMC (RJMCMC) \citep{green1995reversible, green2009reversible}, which is a general class of MCMC algorithms that allows such moves via the concept of dimensionality matching. The full data likelihood function will also be conditional on $M$; without any adjustment to (\ref{lklh}), it will be specified as $f(\dot{\bm{C}}|\bm{\delta},\bm{\Theta}_1,\cdots,\bm{\Theta}_M, \phi,M)=f(\dot{\bm{C}}|\bm{\delta},\bm{\mathcal{T}}, \phi,M)$. 

\subsection{Priors}
We model the number of segments $M$ via a truncated Poisson distribution $M \sim \text{TPoi}(\eta, M_\text{min}, M_\text{max})$
where $\eta$ is a rate parameter and the lower bound is set at $M_\text{\text{min}} = 1$, allowing for no sub-division in the entire data at a minimum (i.e., no change-point exists). Likewise, an upper bound at $M = M_{\text{max}}$ constrains the number of segments to $M \leq M_\text{max}$. A small value of $\eta$ leads to a prior placing high mass on small values of $M$. Selection of $\eta$ has been a challenging problem when implementing RJMCMC for various applications \citep{strait2019automatic}; since our goal is to select a small number of change-points, it is reasonable to treat $\eta$ as a regularization parameter. Therefore, to guard against over-fitting, we may choose the prior on $M$ such that most of its mass is placed very close to zero, to penalize choosing high values of $M$. For our simulation study and real data analysis, we recommend setting $\eta\in(10^{-6},10^{-3})$. 

\subsection{Posterior sampling via RJMCMC}
Performing trans-dimensional moves via RJMCMC requires joint updates across $\bm{\delta}, \bm{\mathcal{T}}$ and $M$. Hence, there is a requirement for obtaining a joint posterior density given by $\pi(\bm{\delta}, \bm{\mathcal{T}}, M| \bm{\dot{C}}, \phi) =  f(\dot{\bm{C}}|\bm{\delta}, \bm{\mathcal{T}}, \phi, M)\pi(\bm{\delta}, \bm{\mathcal{T}}, M)$. Note that the posterior update of $\phi$ remains the same. 

We design RJMCMC by constructing trans-dimensional moves via joint proposal distributions. Consider two models $\mathcal{M}$ and $\mathcal{M}^*$ in states $( \bm{\delta}, \bm{\mathcal{T}}, M)$  \& $(\bm{\delta}^*, \bm{\mathcal{T}}^*, M^*)$ respectively. Without loss of generality, consider $M^* > M$ such that $\bm{\mathcal{T}}^*$ is of a higher dimension and $\sum_{t = 1}^T \delta^*_t = M^* > \sum_{t = 1}^T \delta_t$. One possible trans-dimensional move is updating model $\mathcal{M}$ to $\mathcal{M}^*$ by matching the dimensions of the two models. In order to perform the move, we first generate a random vector $\bm{v}$ from a proposal density function $g_{3M \rightarrow 3M^*}(\cdot)$. Subsequently, $\mathcal{M}$ is updated to $\mathcal{M}^*$ by devising a one-to-one mapping function $h_{3M \rightarrow 3M^*}(\cdot)$ such that $\bm{\mathcal{T}}^* = h_{3M \rightarrow 3M^*}(\bm{\mathcal{T}}, \bm{v})$. Likewise, $\delta^*_t = 1$  for $M^* - M$ available time-points having $\delta_t = 0$ such that $M + (M^* - M) = M^*$. The proposed move from model $\mathcal{M}$ to $\mathcal{M}^*$ is accepted with probability $\min(1, h)$, where:

\begin{equation}
h\quad =\quad \frac{\pi(\bm{\delta}^*, \bm{\mathcal{T}}^*, M^*| \bm{\dot{C}}, \phi)J(\bm{\delta}, M|\bm{\delta}^*, M^*)}{\pi(\bm{\delta}, \bm{\mathcal{T}}, M| \bm{\dot{C}}, \phi)g_{3M \rightarrow 3M^*}(\bm{v})J(\bm{\delta}^*, M^*|\bm{\delta}, M)}\left|\frac{\partial h_{3M \rightarrow 3M^*}(\bm{\mathcal{T}}, \bm{v})}{\partial (\bm{\mathcal{T}}, \bm{v})}\right|.
\end{equation}
The Hastings ratio $h$ can be simplified by considering a certain class of trans-dimensional moves; namely, the {birth} and {death} moves \citep{cappe2003reversible}. The Jacobian term is equal to one for these moves. Reverting to the original notation used for the manual change-point detection algorithm, we can obtain the Hastings ratio as a product of posterior and proposal ratios:
\begin{equation}\label{move}
    h\quad =\quad \frac{\pi(\bm{\delta}^*, \bm{\mathcal{T}}^*, M^*| \bm{\dot{C}}, \phi)}{\pi(\bm{\delta}, \bm{\mathcal{T}}, M| \bm{\dot{C}}, \phi)}\frac{J(\bm{\mathcal{T}}|\bm{\mathcal{T}}^*, M^*)}{J(\bm{\mathcal{T}}^*|\bm{\mathcal{T}}, M)} \frac{J(\bm{\delta}|\bm{\delta}^*, M^*)}{J(\bm{\delta}^*|\bm{\delta}, M)}\frac{J(M|M^*)}{J(M^*|M)}.
\end{equation}

\subsubsection{{Birth move}}
Under the {birth} move, we propose $M^* = M + 1$ such that we move from state $( \bm{\delta}, \bm{\mathcal{T}},M)$ to $(\bm{\delta}^*, \bm{\mathcal{T}}^*,M^*)$, increasing the dimensionality of $\bm{\mathcal{T}} = \{\bm{\Theta}_1, \cdots, \bm{\Theta}_{M}\}$ by three via a one-to-one vector augmenting function $h_{3M \rightarrow 3M^*}(\bm{\mathcal{T}}, \bm{v}) = \left\{\bm{\Theta}_1, \cdots, \bm{\Theta}_m, \right.$ $\left. \exp(\bm{v}),\bm{\Theta}_{m + 1}, \cdots, \bm{\Theta}_{M}\right\} = \left\{\bm{\Theta}^*_1, \cdots, \bm{\Theta}^*_{M + 1}\right\} = \bm{\mathcal{T}}^*$ for a new segment $m^*$ between $m$ and $m + 1$-th segments. In particular, 
a new segment is added such that for some randomly selected $t:z^*_t = m^*$ we have $\delta_t^* = 1$, but $\delta_t = 0$. For the epidemiological parameters of the new segment $m^*$, we propose $\bm{v}$ from a product of independent truncated normal distributions, i.e., $\bm{v} = \ln(\bm{\Theta}^*_{m^*}) \sim \text{TN}(\ln K_m,\sigma_K^2,a_K,b_K)\text{N}(\ln\lambda_m,\sigma_\lambda^2)\text{TN}(\ln p_m,\sigma_p^2,-\infty,0)$, which forms $J(\bm{\mathcal{T}}^*|\bm{\mathcal{T}}, M)$. Regarding other proposal density functions, $J(\bm{\delta}^*|\bm{\delta}, M) = 1/(T - M)$ measures the probability of adding a change-point to an available location. On the other hand, for the reverse move, $J(\bm{\delta}|\bm{\delta}^*, M^*) = 1/(M^* - 1)$ is the probability of deleting a change-point given $M^*$ segments. Also, $J(\bm{\mathcal{T}}|\bm{\mathcal{T}}^*, M^*) = 1$ as a set of parameters is deleted from the parameter space, given the deletion of a segment, with probability one. We also assume that $M$ and $M^*$ are equally likely to be updated such that $J(M|M^*) = J(M^*|M)$. 

\subsubsection{{Death move}}
 A {death} move is considered the inverse of a {birth} move with $M^* = M - 1$ such that a certain segment is dropped (i.e., for some randomly chosen $t:z^*_t = m^*$, ${\delta}_t^* = 0$, but ${\delta}_t = 1$). This reduces the dimensionality $\bm{\mathcal{T}}$ by three. As a set of parameters is deleted from the parameter space, given the deletion of segment $m$, we have $J(\bm{\mathcal{T}}^*|\bm{\mathcal{T}}, M) = 1$. For the reverse move $\mathcal{M} \leftarrow \mathcal{M}^*$, the density $J(\bm{\mathcal{T}}|\bm{\mathcal{T}}^*, M^*)$ proposes a set of parameters for $\mathcal{M}$ on the logarithm scale from a product of independent truncated normal distributions (i.e., $\text{TN}(\ln K_m,\sigma_K^2,a_K,b_K)\text{N}(\ln\lambda_m,\sigma_\lambda^2)\text{TN}(\ln p_m,\sigma_p^2,-\infty,0)$). The proposal probability $J(\bm{\delta}^*|\bm{\delta}, M) = 1/(M - 1)$ measures the probability of deleting a randomly selected change-point given $M$ segments. On the other hand, for the reverse move $\mathcal{M} \leftarrow \mathcal{M}^*$, $J(\bm{\delta}|\bm{\delta}^*, M^*) = 1/(T - M^*)$ is the probability of adding a change-point at an available location. We also assume that $M$ and $M^*$ are equally likely to be updated such that $J(M|M^*) = J(M^*|M)$.

Note that the proposal ratios for {birth} and {death} moves are approximate, subject to the gap constraints that do not allow successive change-points at consecutive time points.



\subsubsection{{Other moves}} Amid the trans-dimensional moves, we also perform a few fixed-dimensional moves to explore the parameter space as efficiently as possible. We include the {global} and {local} swap moves to update $\bm{\delta}$ without perturbing $M$. Subsequently, all epidemiological parameters $\Theta_m \in \bm{\mathcal{T}}$ are updated one at a time via the MCMC sampler. In addition, a {stay} move is also included, which only updates the epidemiological parameters $\Theta_m$ via MCMC for fixed $\bm{\delta}$ and $M$.

At each iteration of the algorithm, the trans-dimensional moves ({birth}, {death}) are performed with probabilities $1/4$ each, while the fixed dimensional moves ({local swap}, {global swap} and {stay}) are performed with probabilities $1/6$ each. For the extreme cases, at $M = 1$, the probability of performing a {death} move is set to zero, while the probability of a {birth} move is increased to $1/2$; likewise, at $M = M_{\text{max}}$, the probability of performing a {birth} move is set to zero, while the probability of a {death} move is increased to $1/2$.

\section{Posterior Inference}\label{sec:post}


The previous two sections discuss MCMC and RJMCMC algorithms that sample parameters under fixed and varying dimensional spaces, respectively. The next step is to compute summary statistics of those MCMC samples for statistical inference. Assume that we run the proposed MCMC sampler for $B$ iterations after burn-in. 

\subsection{Point estimates}
We start by obtaining a point estimate of the change-point indicator $\bm{\delta}$ by analyzing its MCMC samples $\{\bm{\delta}^{(1)},\cdots,\bm{\delta}^{(B)}\}$, where $b$ indexes the MCMC iteration after burn-in. We first consider the \textit{maximum-a-posteriori} (MAP) estimate, defined as:
\begin{equation}
    \bm{\delta}^\text{MAP} \quad = \quad {\arg\max}_b f(\dot{\bm{C}}|\bm{\delta}^{(b)},\bm{\Theta}_1^{(b)},\cdots,\bm{\Theta}_M^{(b)},\phi^{(b)})\pi(\bm{\delta}^{(b)}|M^{(b)}).
\end{equation}
The corresponding $\bm{z}^\text{MAP}$ can be obtained by taking the cumulative sum of $\bm{\delta}^\text{MAP}$. The MAP estimates for the three epidemiological parameters of each sub-epidemic can be obtained in a similar fashion. In addition, we consider another estimate on $\bm{\delta}$ based on posterior probabilities of inclusion (PPIs), each of which is defined as:
 \begin{equation}
     \text{Pr}(\delta_t=1|\bm{\delta}\backslash\delta_t, \bm{\Theta}_1,\cdots,\bm{\Theta}_M,\phi,M) \quad\approx\quad \frac{1}{B}\sum_{b = 1}^B \bm{1}(\delta^{(b)}_t = 1), t = 1,\cdots,T.
 \end{equation}
This allows us to quantify the confidence of considering each time point as a change-point.
 
 \subsection{Interval estimates} 
 \label{sec:interval}
 The estimation of credible intervals for $\bm{\delta}$ is non-trivial as it is updated via swap moves and not proposal densities. We follow \cite{Jiang2020.10.06.20208132} to construct a "credible interval" for each identified change-point through MAP estimation. We utilize its local dependency structure from all MCMC samples of $\bm{\delta}$ that belong to its neighbors. Due to the prior setting described in Section \ref{sec:manual}, if a time point $t$ is selected as a change-point (i.e., $\delta_t=1$) then its nearby time points must not be a change-point. Therefore, the correlation between the MCMC sample vectors $\{\delta_t^{(1)},\cdots,\delta_t^{(B)})$ and $(\delta_{t\pm s}^{(1)},\cdots,\delta_{t\pm s}^{(B)})$ tends to be negative when $s$ is small. We define the credible interval of a change-point as the two ends of all its nearby consecutive time points, for which the MCMC samples of $\bm{\delta}$ are significantly negatively correlated with that of the change-point. This could be done via a one-sided Pearson correlation test with a pre-specified significant level (e.g. $0.05$). Although quantifying uncertainties of change-points is not rigorous, it performs very well in the simulation study and yields reasonable results in real data analysis. The credible intervals for those epidemiological parameters of each sub-epidemic can be easily approximated by their post-burn-in MCMC sample quantiles.

\subsection{Predictive estimates} We forecast the cumulative or new confirmed case number for future time points $t=T+1, \cdots, T+T_f$ via MCMC samples of $\phi$ and the last-segment parameters $K_M$, $p_M$, and $\lambda_M$. For future time points, we sample $\dot{C}_{T+1}^{(b)}$ from $\text{NB}\left(\lambda_m^{(b)} C_{T}^{p_m^{(b)}}\left(1 - \frac{C_{T}}{K_m^{(b)}}\right), \phi^{(b)}\right)$ and sequentially sample $\dot{C}_t,t=T+2,\cdots,T+T_f$ from
\begin{equation}
    \dot{C}_t^{(b)}|z_t=M,\bm{\Theta}_M,\phi \quad\sim\quad \text{NB}\left(\lambda_M^{(b)} \left(C_{t - 1}^{(b)}\right)^{p_M^{(b)}}\left(1 - \frac{C_{t - 1}^{(b)}}{K_M^{(b)}}\right), \phi^{(b)}\right).
\end{equation}
Both short and long-term forecasts can be made by summarizing the $T_f$-by-$B$ matrix of MCMC samples with each entry being $\dot{C}_t^{(b)}$. For instance, the posterior predictive mean of new confirmed cases at time point $t$ can be approximated by $\sum_{b = 1}^{B} \dot{C}_t^{(b)}/B$, while predictive intervals can be approximated by the sample quantiles of the $b$-th column of the matrix.

\section{Simulation Study}\label{sec:sim} 






To measure the change-point detection accuracy of our model, we simulated data using both growth and compartmental modeling frameworks under different noise levels. We further compared BayesSMEG to alternative change-point detection models.

\subsection{Generative models}\label{sec:simusetup}

We first considered $T = 150$ time points divided across $M = 3$ equally-sized segments and a population size of $N = 200,000$. The resulting two change-points were placed at time $t = 52$ and  $103$, respectively. The initial epidemic size was set to $C_0 = 100$. The newly added confirmed cases within each sub-epidemic were then sampled based on GLC, $\dot{C}_t|\cdot \sim \text{NB}\left(\lambda_m C_{t - 1}^{p_m}\left(1 - \frac{C_{t - 1}}{K_m}\right), \phi\right)$ for all $t = 1,\cdots, T$, where the growth rates, final epidemic sizes, and growth scaling factors of the three sub-epidemics were set to $\bm{\lambda} = (0.1, 0.06, 0.08)^\top$, $\bm{K} = (10,000, 9,000, 15,000)^\top$, $\bm{p} = (0.9, 0.85, 0.9)^\top$, respectively. This setup corresponds to the most common scenario observed across all U.S. states (i.e., a higher growth rate in the first sub-epidemic, followed by a lower rate due to government interventions such as business and school closures, and moderately higher rates in the final sub-epidemic characterized by re-openings). We considered high and low dispersion scenarios attributed by $\phi=10$ and $100$, respectively. We repeated the above steps to generate $50$ independent simulated datasets for each setting in terms of $\phi$.

To demonstrate the versatility of our method on other epidemiological frameworks, we also generated data that does not favor BayesSMEG. In particular, we followed a generative scheme based on the segmented SIR model proposed in \cite{Jiang2020.10.06.20208132}; see Section S3.1 in the supplement for more details. A short summary of the simulated SIR data is that $T = 120$ time points were divided into $M=4$ equal segments, given the resulting three change-points set at $t = 31, 61, 91$-st time points. The population size was $N = 1,000,000$.

\subsection{Competing methods}
We evaluated the performance of BayesSMEG and compared it to three alternative change-point detection methods. The first was a Bayesian change-point detection model \citep{barry1993bayesian} that uses a product partition modeling approach to conduct Bayesian inference on change-points. \cite{erdman2007bcp} relaxed certain modeling assumptions and developed \texttt{bcp}, an \texttt{R} package. In particular, we performed $100,000$ MCMC runs using \texttt{bcp}, and discarded the first half as burn-in. Subsequently, the posterior probabilities of being change-points $\pi({\delta}_t=1|\cdot)$'s were generated and visualized. To create a fair comparison in clustering evaluation, the first $M$ time points with the highest posterior probabilities were considered change-points. The second alternative was a deterministic approach of fitting regression models with unknown break-points \citep{muggeo2003estimating}. The related \texttt{R} package, \texttt{seg} \citep{muggeo2008segmented}, returns an estimated segmentation vector $\hat{\bm{z}}$ for a pre-specified $M$. The response and predictor variables were $\ln(\dot{\bm{C}})$ and time indices $(1,\cdots,T)^\top$, respectively. The third alternative was the recently developed BayesSMILES \citep{Jiang2020.10.06.20208132}, which performs automatic change-point detection via a Bayesian Poisson segmented regression model with SIR dynamics.

\subsection{Evaluation criteria} The goal of the simulation study was to evaluate and compare the performance of change-point detection methods under different settings. This was accomplished by measuring discrepancies between true and estimated change-point segmentation vectors $\bm{z}$ and $\hat{\bm{z}}$, which could be immediately derived from their corresponding change-point indicator vector $\bm{\delta}$ and $\hat{\bm{\delta}}$, respectively. Segmentation can be thought of as a special case of a clustering problem. Thus, we employed common clustering performance metrics, including the adjusted Rand index (ARI), mutual information (MI), and normalized variation of information (NVI). As change-point detection can be viewed as a classification problem, we also chose the F-measure as a supplementary metric. For rigorous definitions of the above metrics in the context of this paper, please refer to Section S3.2 in the supplement.

\subsection{Results}
\subsubsection{Estimation of change-point number and locations}
For the implementation of BayesSMEG, we followed the hyper-parameter recommendation detailed in Section \ref{sec:prior} and set $\eta = 10^{-4}$ or $10^{-3}$, $M_{\text{max}} = 50$. We set $\rho = 0.3$, assuming that at most $30\%$ of the entire population of a particular region would be infected over the epidemic. We performed $100,000$ MCMC runs, discarding the first half as burn-in ($B=50,000$). Algorithm settings to control MCMC step-size were set to $\sigma_{\phi} = \sigma_{K} = 1$, $\sigma_{\lambda} = \sigma_{p} = 0.1$.

\begin{figure}[!h]
    \centering
    \includegraphics[width = 1.0\linewidth]{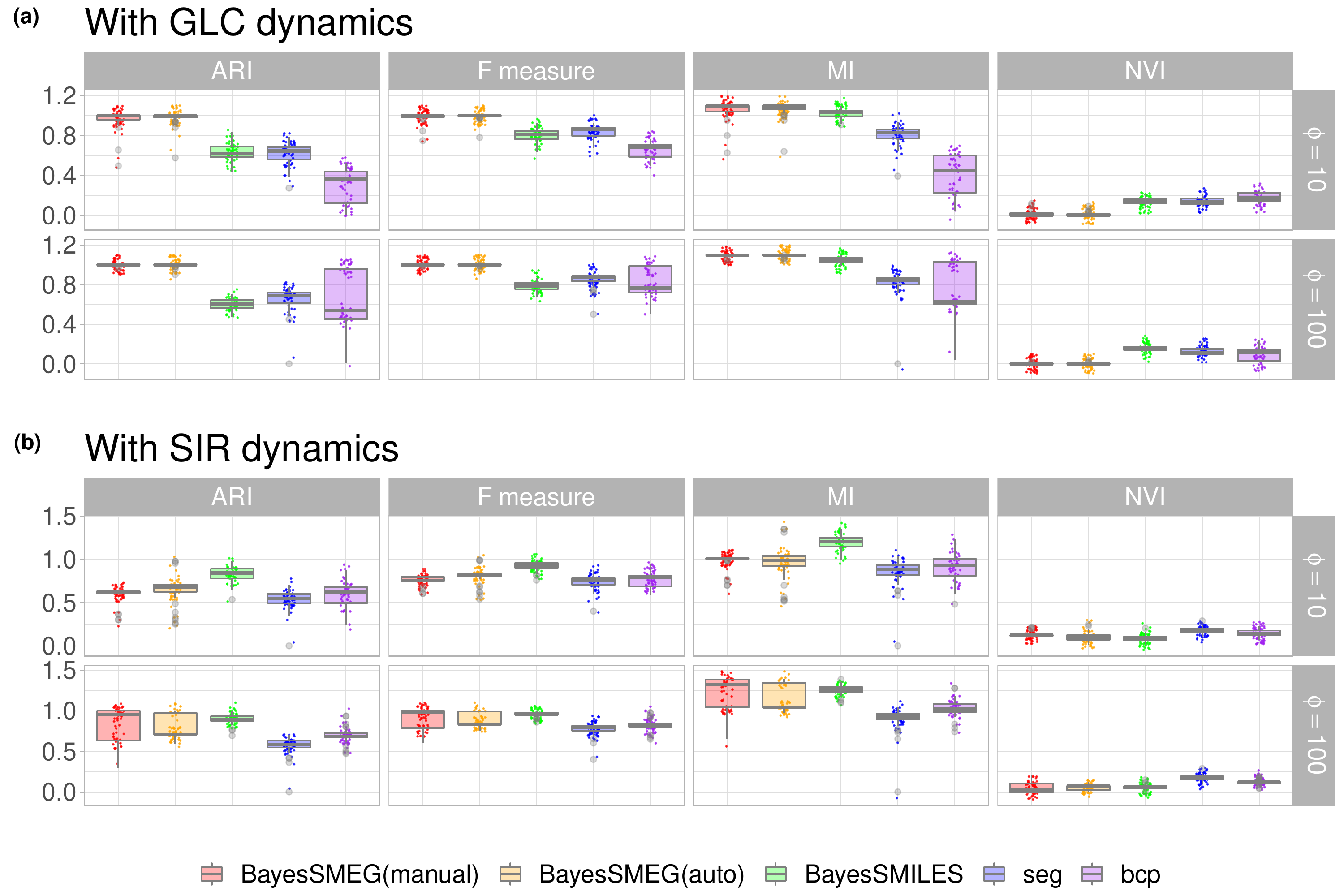}
    \caption{Simulation study: Boxplots of change-point detection accuracy in terms of adjusted Rand index (ARI), F-measure, mutual information (MI), and normalized variation of information (NVI) by BayesSMEG and alternative methods, under different scenarios in terms of dispersion parameter $\phi$ and generative schemes: (a) Generalized logistic curve (GLC) and (b) Susceptible-infected-removed (SIR) dynamics.}
    \label{fig:fig1}
\end{figure}

Figure \ref{fig:fig1} depicts the computed ARI, F-measure, MI, and NVI by all methods over $50$ replicated simulated datasets for each setting in terms of $\phi$ and generative scheme. It is observed that BayesSMEG was the most accurate change-point detection algorithm, irrespective of the two dispersion settings, when the simulated data were generated with growth dynamics. In contrast, when data were generated based on the SIR model, BayesSMILES generally performed better. It is also noted that \texttt{bcp} performed much worse under highly variable simulations ($\phi=10$), as it failed to distinguish between deviations by chance and true structural deviations of epidemiological importance. 



\begin{figure}[!h]
    \centering
    \includegraphics[width = 1.0\linewidth]{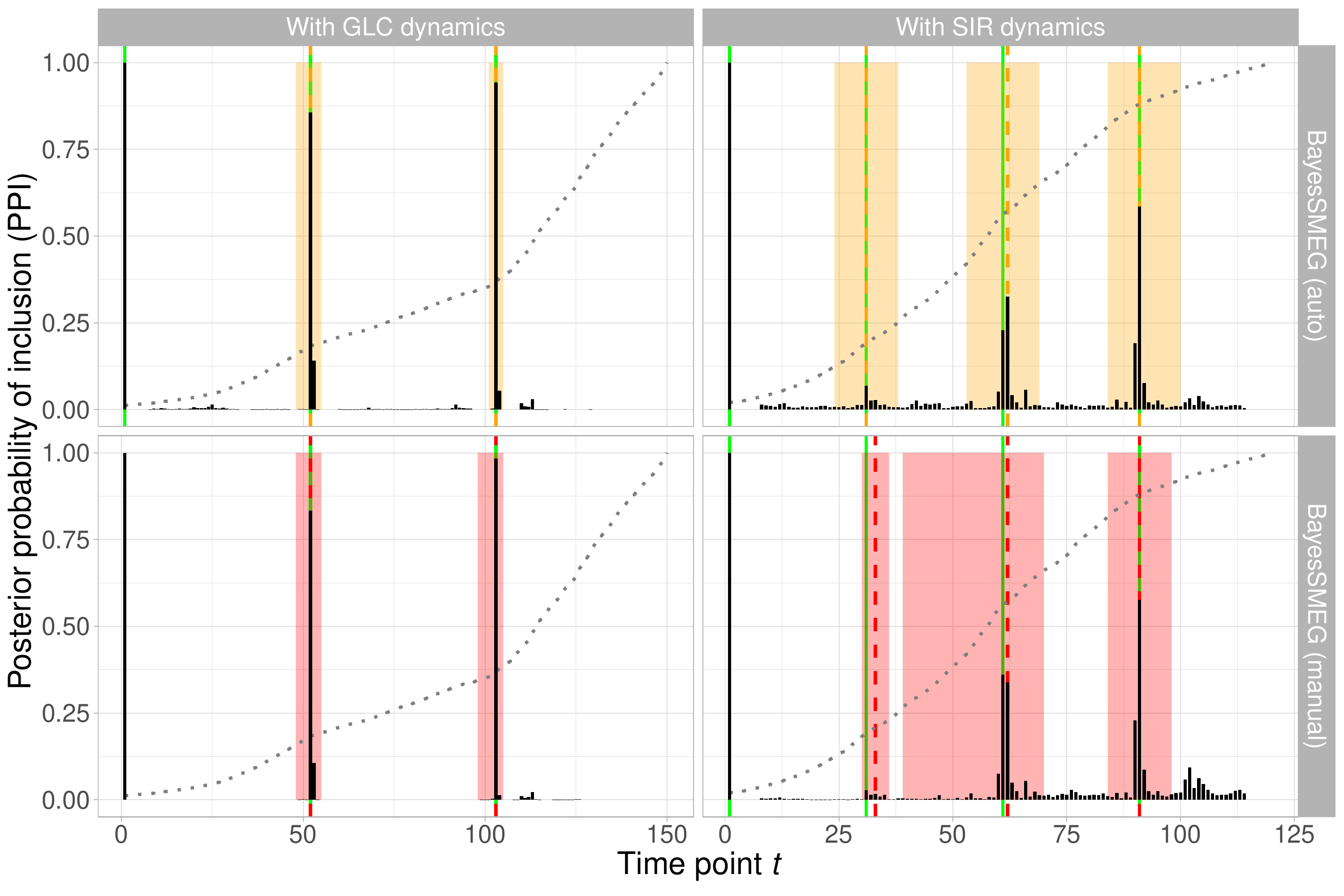}
    \caption{Simulation study: BayesSMEG correctly identified all change-points from the simulated cumulative confirmed cases (black dots; rescaled to a range of $[0,1]$) generated by the two generative schemes: the generalized logistic curve (GLC) and susceptible-infected-removed (SIR) dynamics. The black solid lines indicate the posterior probabilities of inclusion (PPIs), $\pi(\delta_t|\cdot)$; the green dashed lines indicate the true change-points; the red dashed lines indicate the identified change-points based on the \textit{maximum-a-posteriori} (MAP) estimates, $\{t:\delta_t^\text{MAP}=1\}$; the shaded orange/red regions indicate the credible intervals for those identified change-points.}
    \label{fig:fig2}
\end{figure}

Figure \ref{fig:fig2} depicts the change-point detection performance of BayesSMEG on a randomly selected simulated dataset generated with growth and SIR dynamics, respectively. Both versions of BayesSMEG, with or without fixing $M$, were evaluated. For growth simulated data, both models retrieved the true change-points with accuracy having PPIs between $80\%$ to $90\%$ with narrow credible intervals. However, for the SIR simulation setting, the algorithms measured the posterior probabilities of true change-points at $5\%$--$10\%$, $25\%$--$35\%$, and $60\%$, respectively. We also observed that the MAP estimate is robust to lower PPIs for the first change-point at $t = 31$ as it collectively maximized the joint log posterior density irrespective of a lower PPI. The credible intervals of change-points appear to be wider compared to the growth simulated data, to account for a higher uncertainty.


\subsubsection{Estimation of final epidemic sizes}
\begin{figure}[!h]
    \centering
    \includegraphics[width = 1.0\linewidth]{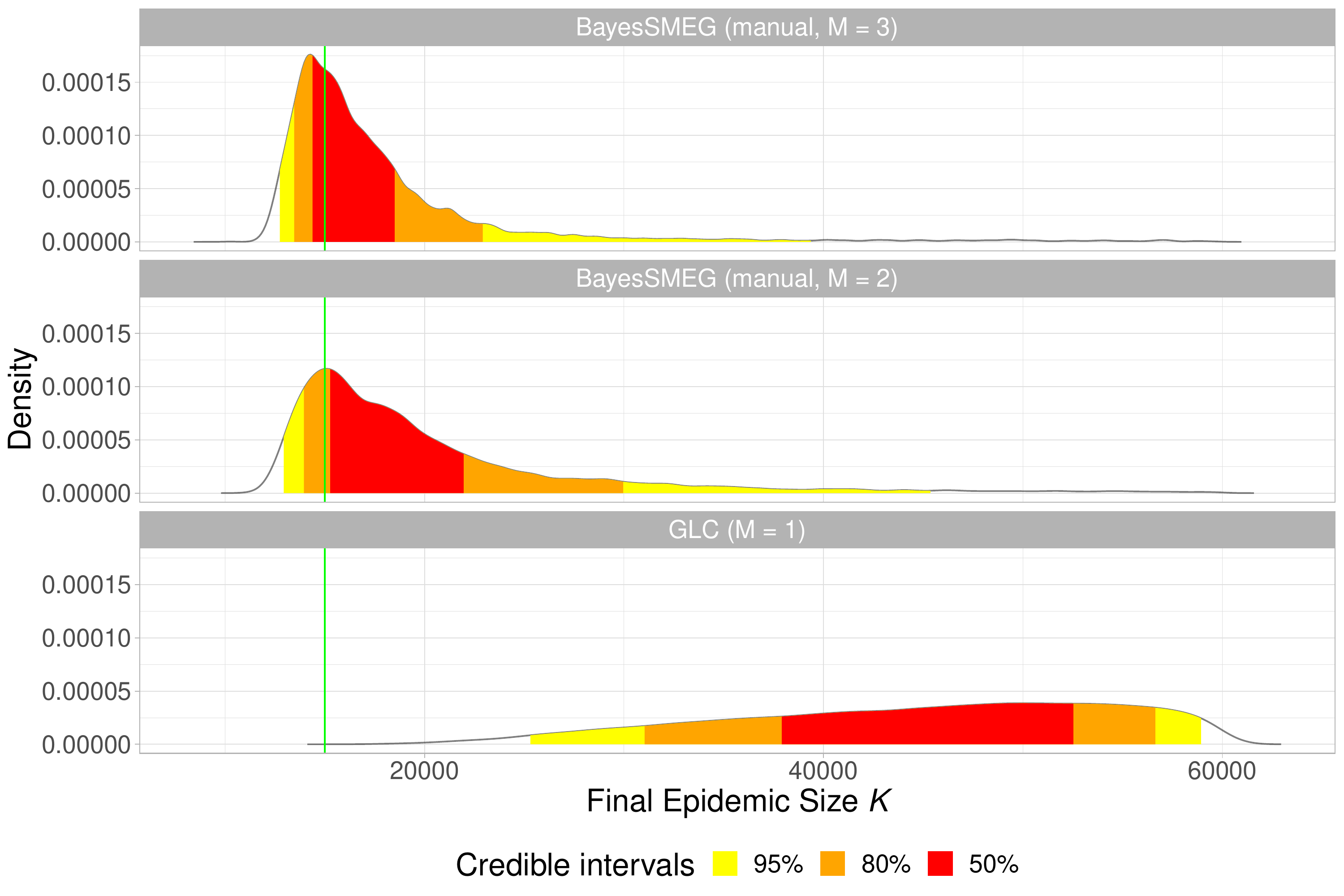}
    \caption{Simulation study: Posterior distributions of the final epidemic size $K_M$ by (a) and (b) BayesSMEG (manual, $M=2$ and $3$) and (c) The generalized logistic curve (GLC) model. Green solid lines indicate the true value of final epidemic size ($K_3 = 15000$); the boundaries of red, orange, and yellow regions indicate $50\%$, $80\%$, and $95\%$ credible intervals.}
    \label{fig:fig3}
\end{figure}

Apart from change-point detection, BayesSMEG utilizes attributes from a GLC for $M$ sub-epidemics in a piece-wise manner. One of its key features includes the estimation of the final epidemic size $K$ for each sub-epidemic. In Figure \ref{fig:fig3}, we fitted BayesSMEG with $M = 3$ and $2$ and GLC ($M=1$) on the simulated growth data generated with growth dynamics. For BayesSMEG, only the posterior densities of $K_M$ for the respective final sub-epidemic were considered as they were supposed to be closest to the true $K$. BayesSMEG with $M = 3$ was able to capture the true value of $K_3=15,000$ within $95\%$, $80\%$, and $50\%$ credible intervals being $[12,728,\, 30,510]$, $[13,541,\, 22,190]$, and $[14,498,\, 18,492]$, respectively. On the contrary, a simple GLC with a $95\%$ credible interval of $[26,089,\, 58,999]$ completely missed the true estimate of $K$ and heavily over-estimated the final epidemic size. BayesSMEG with $M = 2$ also captured the true value of final epidemic size within a $95\%$ credible interval of $[12,879,\, 47,958]$. The MAP estimates of $K$ of BayesSMEG were fairly close to the true value (i.e., $\hat{K}^\text{MAP} = 14,539$, and $14724$ for the two cases, respectively). 


\section{Analysis of COVID-19 Data}\label{sec:res}

In this section, we fit BayesSMEG to the daily numbers of confirmed COVID-19 cases in California and New York to estimate and interpret change-points. We considered the first $T=500$ days (between March 9, 2020, and July 21, 2021) after the cumulative confirmed cases rose above $100$. We followed the hyper-parameter recommendation detailed in Section \ref{sec:prior} and also assumed that $\rho = 0.3$, indicating that at most, $30\%$ of the entire population of California or New York could be infected by COVID-19. Furthermore, we set $\eta = 10^{-5}$ and $M_\text{max} = 50$ to infer $M$ using an automatic change-point detection algorithm via RJMCMC. Algorithm settings to control MCMC step-size were set to $\sigma_{\phi} = \sigma_{K} = 1$, $\sigma_{\lambda} = \sigma_{p} = 0.1$.

\subsection{Estimation of change-point number and locations} 

\begin{figure}[h]
    \centering
    \includegraphics[width = 1.0\linewidth]{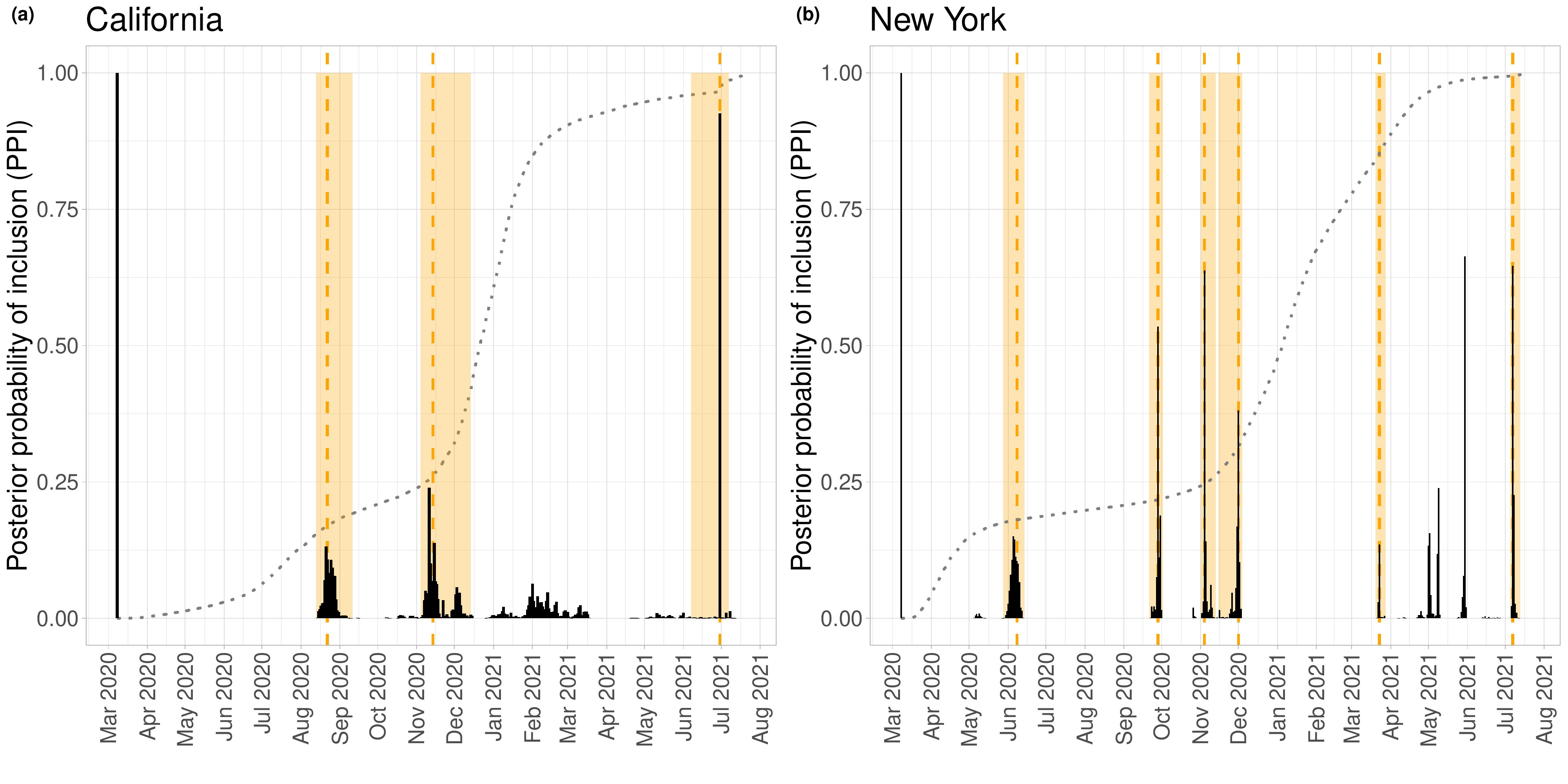}
    \caption{Real data analysis: BayesSMEG (auto) identified three and six change-points from the cumulative confirmed COVID-19 cases (black dots; rescaled to a range of $[0,1]$) in (a) California and (b) New York, respectively. Black solid lines indicate the posterior probabilities of inclusion (PPIs), $\pi(\delta_t|\cdot)$; red dashed lines indicate the identified change-points based on the \textit{maximum-a-posteriori} (MAP) estimates, $\{t:\delta_t^\text{MAP}=1\}$; shaded orange regions indicate the credible intervals for those identified change-points.}
    \label{fig:fig4}
\end{figure}

 On March 11, 2020, the World Health Organization (WHO) declared COVID-19 a pandemic. Soon after, the governor of California responded by restricting public gatherings to $250$ people in concentrated areas and declaring school closures. By March 17, several counties had imposed shelter-in-place orders to "flatten the curve". Due to prompt government intervention, the growth in cases was much more gradual than expected; bolstered by these positive results, the government planned on eventually loosening the restrictions. However, cases, deaths, and hospitalizations rose rapidly from May to July, breaking the daily case record on July 20. By August 12, cases and hospitalizations had dipped, declining almost $20\%$ over two weeks. As shown in Figure \ref{fig:fig4}(a), BayesSMEG estimated the first change-point as August 22, 2020, marking the end of the first sub-epidemic. This segment was characterized by a more gradual growth in cases after restrictions were imposed, an exponential rise when restrictions were loosened, and an eventual decay in cases. On September 16, 2020, when case numbers hit a record low, the state began loosening business restrictions; death counts continued to plummet by $57\%$ over the past two weeks beforehand. Cases continued to decline in September and October. However, the first week of November saw a continuous surge in cases. BayesSMEG estimated November 11, 2020 as the second change-point marking the beginning of another sub-epidemic cycle. Large spikes in cases were reported in the month of December, with the state reaching maximum ICU capacity on Christmas. The first COVID-19 vaccine was deployed on December 14, and during the second week of January 2021, California began relaxing vaccine eligibility requirements, allowing more individuals to receive it. By the first week of March 2021, roughly seven million individuals had received their first dose of the vaccine and nearly half of them had also received their second dose. Coinciding with rising vaccination rates, cases and hospitalization rates were recorded to be low and stable. 
 However, after California re-opened on June 15, 2021, another surge of cases was observed due to the rise of the Delta variant. Case numbers doubled in July compared to the past two weeks beforehand, marking the beginning of another wave as detected by BayesSMEG. 

The state of New York detected its first COVID-19 case on March 1, 2020. Subsequently, the governor declared a state of emergency and restricted gatherings to $500$ people. Through March and April, the governor ordered several public health measures, including closures of public schools, bars, and restaurant, and required people to wear masks in public places. Towards the end of May 2020, reported cases appeared not to increase as drastically as previously observed; this could be attributed to stringent government interventions that resulted in reduced disease transmission. Figure \ref{fig:fig4}(b) shows that the first change-point was detected on June 9, 2020 by BayesSMEG, coinciding with phase 1 of the reopening of New York City on June 8, followed by phase 2 on June 22, 2020. This marked the end of the sub-epidemic as observed in New York. By July 19, the government had also begun phase 3 and phase 4. Come September, gyms and malls were allowed to reopen at limited capacity. Elementary, middle, and high school students in New York City were allowed to attend in-person classes by October 1. BayesSMEG estimated the second change-point as October 1, 2020, marking the beginning of another surge in cases and increased transmission rates. Over the following weeks, New York reached the grim milestone of over half a million confirmed COVID-19 cases. Other significant change-points coincided with the dates of new COVID-19 restrictions on limited public gatherings with up to ten people in private homes, the reopening of elementary schools and high schools on November 19, 2020, December 7, 2020, and March 24, 2021, respectively. The final change-point was estimated to be July 8, 2021, marking another surge in cases due to the emergence of the Delta variant, similar to what was observed in California. 

\subsection{Long-term trend forecasting}
We demonstrated the efficacy of BayesSMEG in estimating final epidemic size on simulated data; however, extending this problem to COVID-19 data is challenging, considering the countless waves caused by new mutations and other unknown factors. Instead, we choose to demonstrate BayesSMEG's superiority in long-term forecasting compared to the GLC and standard SIR models. 

\begin{figure}[h]
    \centering
    \includegraphics[width = 1.0\linewidth]{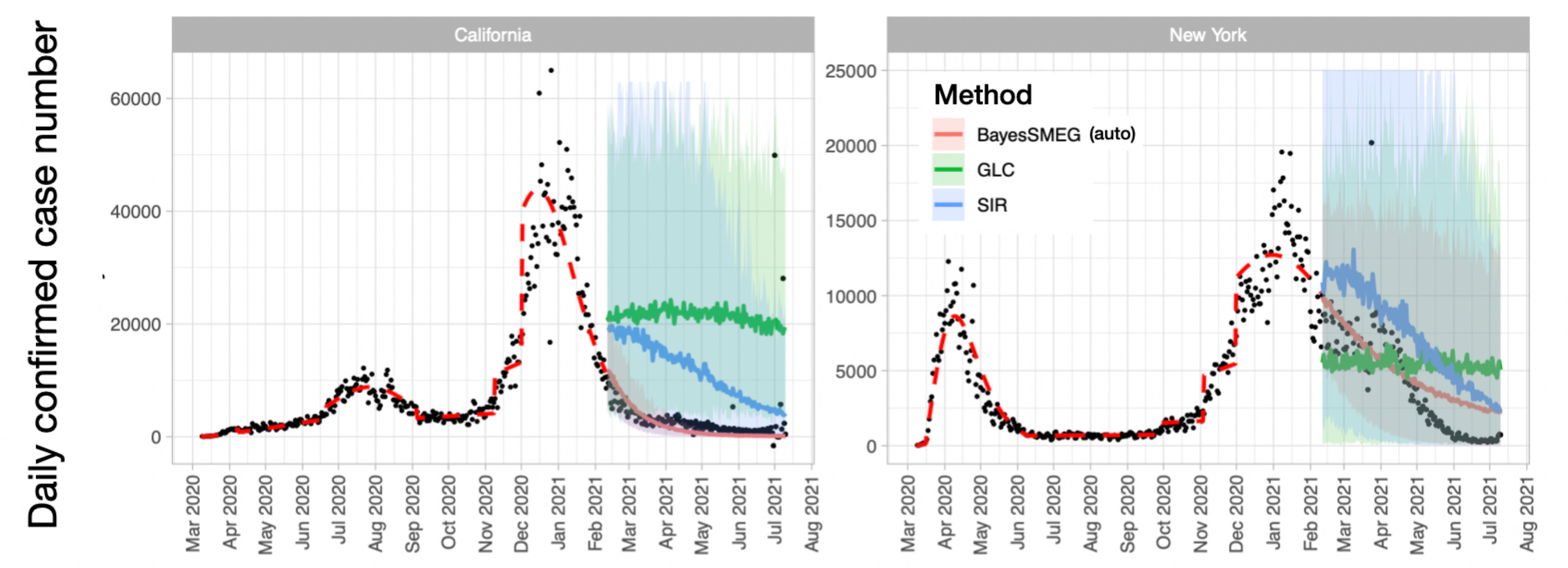}
    \caption{Real data analysis: Fitted curves and predictive trends for the daily confirmed COVID-19 cases (black dots) in (a) California and (b) New York by BayesSMEG (auto), the generalized logistic curve (GLC) model, and the standard susceptible-infected-removed (SIR) model. Red dashed lines are the fitted curves by BayesSMEG (auto); red, green, and blue solid lines are the daily predictive means by BayesSMEG (auto), GLC, and SIR; shaded red, green, and blue regions indicate the daily predictive intervals by BayesSMEG (auto), GLC, and SIR.}
    \label{fig5}
\end{figure}

\begin{figure}[h]
    \centering
    \includegraphics[width = 1.0\linewidth]{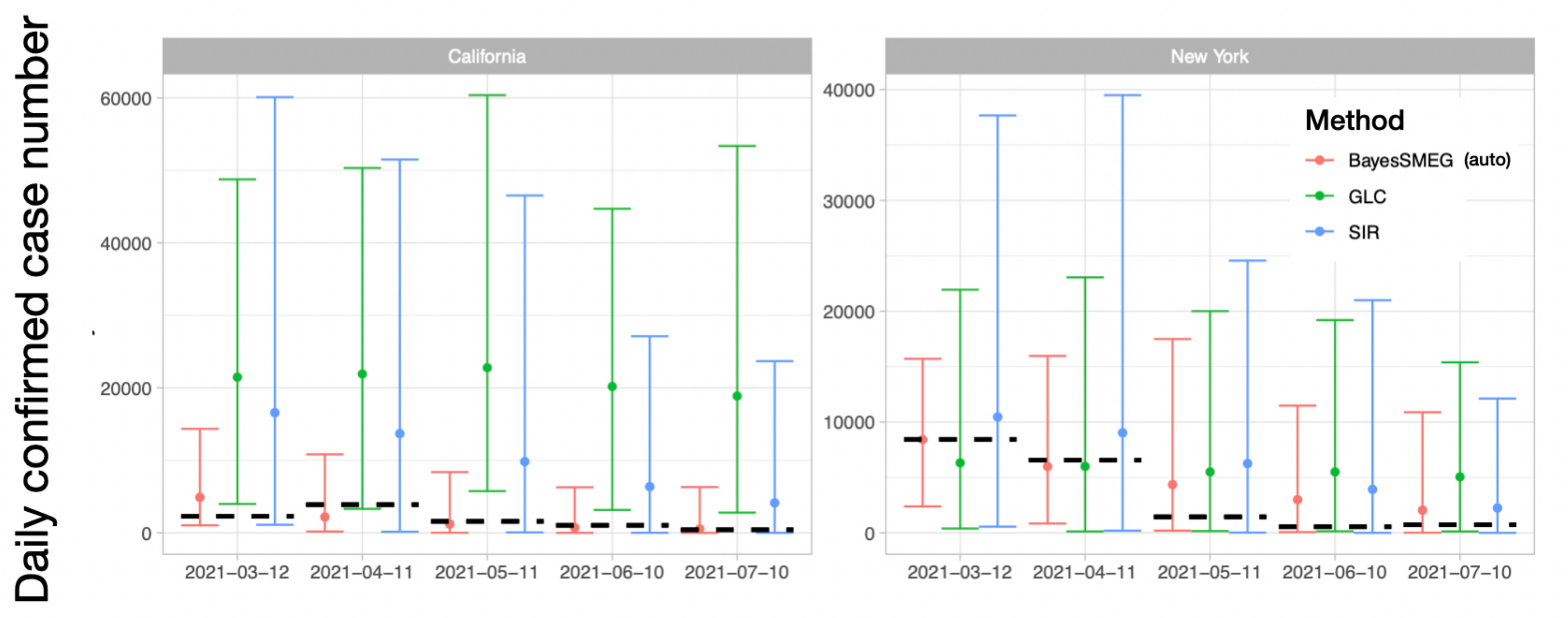}
    \caption{Real data analysis: Observed daily confirmed COVID-19 cases (black solid lines) at five equally spaced time points in (a) California and (b) New York, with corresponding predictive means and intervals by BayesSMEG (auto) in red, the generalized logistic curve (GLC) model in green, and the standard susceptible-infected-removed (SIR) model in blue.}
    \label{fig6}
\end{figure}

All models were fit to daily confirmed COVID-19 cases from March 8, 2020 to February 10, 2021 ($T=500$). Their predictive performance was evaluated on the next $T_f=150$ days from February 11, 2021 to July 10, 2021. As shown in Figure \ref{fig5}(a), the GLC and SIR models clearly overestimated daily cases. In addition, the estimates generated by SIR and GLC were highly uncertain due to their extremely wide predictive intervals. On the other hand, BayesSMEG obtained precise predictive mean and interval estimates close to the true case numbers. As a result, BayesSMEG successfully captured the growth dynamics of the final phase of the sub-epidemic. The other models struggled to detect the steep drop in California, compared to a gradual drop in New York. In Figure \ref{fig5}(b), we closely examined long-term forecasts with consecutive gaps of $30$ days. BayesSMEG always contained the true number of cases within the bounds for both states. In contrast, the other models sometimes missed the mark by a considerable margin and had extremely wide interval estimates, affecting their credibility. 

Other techniques of forecasting accuracy include a comparison of mean absolute percentage error (MAPE) metrics across different models for a particular region; however, this metric is undefined if there are zero cases at any given time point. To circumvent this drawback, we computed adjusted MAPE (AMAPE) given by:
\begin{equation}
    \text{AMAPE}(\hat{\bm{\dot{C}}}, \bm{\dot{C}}) = \frac{1}{T_f}\sum_{t = T + 1}^{T+T_f}\left|1 - \frac{\hat{\dot{C}}_{t}}{\dot{C}_t + \bm{1}(\dot{C}_t = 0)}\right|.
\end{equation}
where $\dot{C}_t$ and $C_t$ are the predictive mean and true case number at time point $t$, respectively, and $\bm{1}(\cdot)$ denotes the indicator function, which serves as the adjustment factor to correct for true zero cases by adding one to the denominator in the presence of a zero. The numerical summary is shown in Table \ref{table1}, showing that the AMAPE estimates obtained by BayesSMEG are significantly smaller than those obtained by GLC and SIR.

\begin{table}[h!]
\caption{Real data analysis: Adjusted mean absolute percentage errors (AMAPEs) of $T_f=150$-day forecasts of daily confirmed COVID-19 cases in California and New York by BayesSMEG (auto), the generalized logistic curve (GLC) model, and the standard susceptible-infected-removed (SIR) model.}
\begin{center}
\begin{tabular}{r|rr}
 & California & New York\\\hline
BayesSMEG (auto) & $8.4$ & $0.9$ \\
GLC & $395.6$ & $2.0$\\
SIR & $89.3$ & $1.2$\\
\end{tabular}
\end{center}
\label{table1}
\end{table}

\subsection{Spatial analysis of all U.S. states based on change-point locations}
We also performed an exploratory analysis of spatial patterns across all $50$ U.S. states in terms of change-point locations. The intent was to allow for an overall comparison of COVID-19 dynamics across all states and the potential discovery of interesting spatial patterns. This analysis was performed on the dataset of daily confirmed COVID-19 cases reported across $17$ months from March 2020 to July 2021. We first fit BayesSMEG to estimate the number and location of change-points for each state, setting $\eta = 10^{-6}$ to allow for fewer change-points. Next, we set up $17$-dimensional binary vectors for each state having one for at least one change-point in a given month and zero for no change-points throughout the month. We then performed a hierarchical clustering on the binary vectors using complete linkage and a binary distance metric to group states with similar change-point locations. 

Figure \ref{fig6}(a) summarizes the hierarchical clustering results, and Figure \ref{fig6}(b) colors all $50$ states based on the assigned clusters. We found that adjacent states tended to be in the same cluster. For instance, North Carolina, South Carolina, Georgia, and Alabama were all within cluster 1, because these states had common change-points in the months of April and August 2020, according to Figure \ref{fig6}(a). Other examples include Washington and Oregon in cluster 4 and Kansas and Oklahoma in cluster 5. Additionally, we observed an interesting change-point pattern in cluster 3 with Nebraska, Colorado, New Mexico, Michigan, Ohio, Indiana, and Minnesota. Most of these states have shared change-points in July 2020 and either March or April 2021. In July 2020, there was a surge in cases across these states. Meanwhile, Nebraska initiated phase 4 reopening for some counties, removing capacity limits on restaurants, bars, and childcare facilities. However, the governor of Colorado ordered all bars and nightclubs in the state closed close due to a surge in cases. Similarly, New Mexico also rolled back its decision on state reopening due to increasing cases. The second shared change-point was observed in the months of March and April 2021. During this time, most of these states allowed all residents aged $50$ and older to be eligible for vaccinations. 



\begin{figure}[t]
    \centering
    \includegraphics[width = 1.0\linewidth]{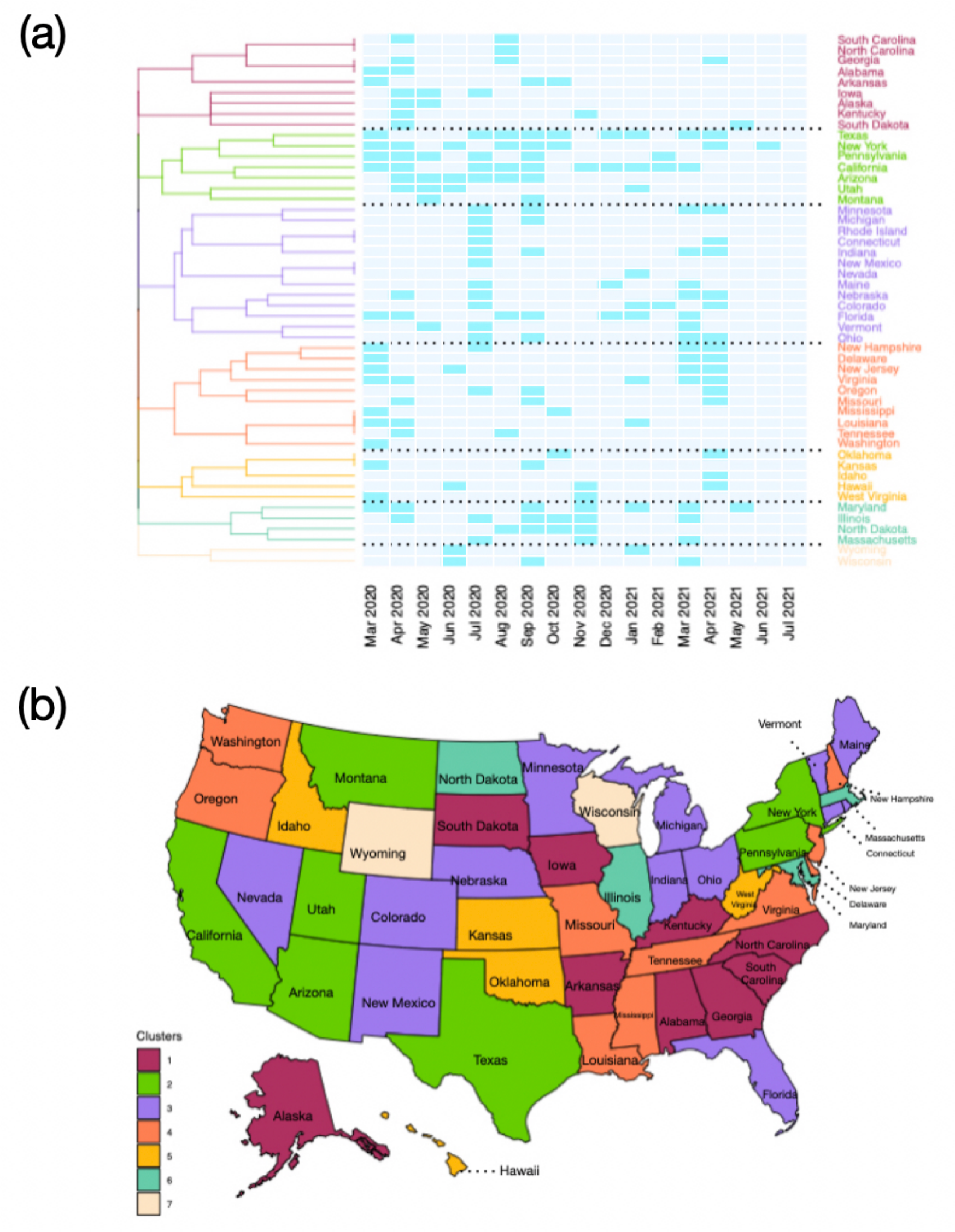}
    \caption{Real data analysis: (a) A heatmap of change-point locations across all U.S. states and the resulting dendrogram from the hierarchical clustering of states based on their change-point locations. (b) A map of the U.S. where each state is colored according to the hierarchical clustering results.}
\label{fig7}
\end{figure}

\section{Conclusion}\label{sec:conc}
In this paper, we proposed BayesSMEG, a Bayesian segmentation modeling framework of epidemic growth. BayesSMEG allows for analyzing epidemiological longitudinal data and understanding infectious disease dynamics with interpretable parameters. The model jointly estimates parameters of epidemiological interest, including sub-epidemic specific growth rates, final epidemic sizes, and the number and locations of change-points. 
The collective estimation of parameters and change-point locations also allows us to perform short and long-term forecasting based on the last sub-epidemic model parameters. Furthermore, the Bayesian setup provides a flexible framework that accounts for measurement error, incorporates prior information if available, and generates sampling distributions of model parameters to quantify their uncertainties. We demonstrated the strength of our model on both simulated and real data. Lastly, we performed a spatial analysis to cluster all $50$ U.S. states into groups of similar disease dynamics based on their change-point locations. Nonetheless, the results of the change-point analysis presented in this paper should be interpreted with some degree of caution as the COVID-19 dataset suffers from accuracy concerns caused by testing limitations, underreporting, and asymptomatic cases. 

For future work, we are working on extending this model to online change-point detection, which would allow the detection and inference of change-points in future unobserved data. In addition, this change-point detection modeling structure has myriad other applications, including finance, stock pricing, and longitudinal biomedical data. 






\begin{acks}[Acknowledgments]
The authors would like to thank Kevin W. Jin for helping us proofread the manuscript, and thank the anonymous referees, the associate editor, and the editor for their constructive comments that improved the quality of this paper.
\end{acks}

\begin{funding}
This work has been supported by the National Institutes of Health [1R01DK131267, 1R01GM141519, 1R01MH128085] and the National Science Foundation [1918854, 2113674, 2210912]. The funding bodies had no role in the design, collection, analysis, or interpretation of data in this study.
\end{funding}


\begin{supplement}[id=suppA]
	\sname{Supplement A}
	\stitle{Software}
	\slink[url]{https://github.com/tejasvbedi95/BayesSMEG.git}
	\sdescription{All software developed in this work is available as \texttt{R} and \texttt{C++} code on GitHub.  We also supply the associated \texttt{R} scripts to reproduce all figures: \url{https://github.com/tejasvbedi95/BayesSMEG.git} }
\end{supplement}

\begin{supplement}[id=suppB]
	\sname{Supplement B}
	\stitle{Supplementary Material}
	\sdatatype{supplment.pdf}
	\sdescription{We provide two tables to summarize the key notation used throughout the paper and additional materials to support Section 3.}
\end{supplement}


\bibliographystyle{imsart-nameyear} 
\bibliography{refs.bib}       

\end{document}


\begin{frontmatter}
\title{\bf Supplementary Materials for\\
	``Bayesian Segmentation Modeling of Epidemic Growth"}

\begin{aug}
\end{aug}

\end{frontmatter}

\beginsupplement
\begin{table}[h!]
	\centering
	\caption{\textbf{Table S1}: Key notation of the proposed BayesSMEG model} 
	\renewcommand{\arraystretch}{1.25}
	\begin{tabular}{p{0.25cm}p{4.25cm}p{3.25cm}p{6.25cm}}
		\bottomrule
		& Notation & Support & Definition \\ \midrule
		\multirow{4}{*}{\rotatebox[origin=c]{90}{Data}}	
		&$N$ & $N \in\mathbb{N}^+$ &The total population size\\
  &$C_0$ & $C_0 \in\{1,\cdots,N\}$ &The initial confirmed case number\\
		&$\bm{C} = \big[ C_t \big]_{T \times 1}$ & $C_t \in\mathbb{N}, C_{t-1}\le C_t\le N$ &The cumulative confirmed case numbers \\
		&$\bm{\dot{C}} = \big[ \dot{C}_t \big]_{T \times 1}$ & $\dot{C}_t = {C}_t - {C}_{t - 1}$ &The newly added confirmed case numbers \\\hline

	\multirow{17}{*}{\rotatebox[origin=c]{90}{Parameters}}
 &$M$ &$M \in \mathbb{N}^+$ &The number of change points\\
		&$\bm{\delta}=\big[\delta_t\big]_{T\times1} $ &${\delta}_t\in \{0, 1\}$  &The change-point indicator vector \\
		&$\bm{z}=\big[z_t\big]_{T\times1} $ &$z_t\in\{1,\cdots,M\}$  &The change-point segmentation vector \\
		&$\bm{\lambda} = \big[\lambda_m\big]_{M\times1}$ &$\lambda_m\in \mathbb{R}^+$ &The early growth rate vector\\
		&$\bm{K} = \big[K_m\big]_{M\times1}$ &$K_m\in \mathbb{R}^+$ &The final epidemic size vector\\
		&$\bm{p} = \big[p_m\big]_{M\times1}$ &$p_m\in [0,1]$ &The growth scaling factor vector\\
		&$\bm{\Theta}=\big[\bm{\Theta}_m\big]_{M\times1}$ &$\bm{\Theta}_m=\{\lambda_m, K_m, p_m\}$ &The superset of epidemiological parameters\\
		&$\phi$ &$\phi \in \mathbb{R}^+$ &The dispersion parameter\\
		\\
		&$a_\lambda,b_\lambda$ & $a_\lambda,b_\lambda\in\mathbb{R}^+$ &The hyper-parameters for $\lambda_m$ \\
		&$a_K,b_K$ & $a_K,b_K\in\mathbb{R}^+$ &The hyper-parameters for $K_m$ \\
		&$a_p,b_p$ & $a_p,b_p\in\mathbb{R}^+$ &The hyper-parameters for $p_m$\\
		&$a_\phi,b_\phi$ & $a_\phi,b_\phi\in\mathbb{R}^+$ &The hyper-parameters for $\phi$\\
  &$\eta$ & $\eta\in\mathbb{R}^+$ &The hyper-parameter for $M$\\
  &$Q$ & $Q\in\{1,\cdots,T\}$ & The minimal number of time points in a segment\\
  &$\rho$ & $\rho\in(0,1]$ & The proportion of population likely to get infected\\
  &$M_\text{max}$ & $M_\text{max}\in\mathbb{N}^+$ & The maximum number of change points\\\hline
		\multirow{3}{*}{\rotatebox[origin=c]{90}{Others}}
		& $\lceil \cdot \rceil$ &   & The ceiling function \\
		& $\bm{1}(\cdot)$ &   & The indicator function \\
		& $|\cdot|$ & & The absolute value function\\
		\bottomrule
	\end{tabular}
	\label{poisson_para_tab}
\end{table}

\begin{table}[h!]
	\centering
	\caption{\textbf{Table S2}: Additional key notations of the stochastic SIR model}
	\renewcommand{\arraystretch}{1.25}
	\begin{tabular}{p{0.25cm}p{4.25cm}p{3.25cm}p{6.25cm}}
		\bottomrule
		& Notation & Support & Definition \\ \midrule
		\multirow{3}{*}{\rotatebox[origin=c]{90}{Data}}	
		&$\bm{S} = \big[ S_t \big]_{T \times 1}$ & $S_t \in\mathbb{N}$ &The susceptible case numbers \\
		&$\bm{I} = \big[ I_t \big]_{T \times 1}$ & $I_t \in\mathbb{N}$ &The infected case numbers \\
		&$\bm{R} = \big[ R_t \big]_{T \times 1}$ & $R_t \in\mathbb{N}$ &The removed (recovered and dead) case numbers \\\hline

		\multirow{5}{*}{\rotatebox[origin=c]{90}{Parameters}}
		&	$\bm{\beta} ~= \big[ \beta_m\big]_{M \times 1}$ & $\beta_m\in\mathbb{R}^+$ &The disease transmission rates for each segment\\
		&	$\gamma$ & $\gamma\in\mathbb{R}^+$ &The pre-specified constant removal rate\\
		&	$\bm{\mathcal{R}}_0 = \big[ \mathcal{R}_{0m} \big]_{M \times 1}$ & $\mathcal{R}_{0m}\in\mathbb{R}^+$ &The basic reproduction numbers for each segment\\
		&	$\phi_S$ &$\phi_S\in\mathbb{R}^+$ &The dispersion parameter for susceptible cases\\
  &	$\phi_R$ &$\phi_R\in\mathbb{R}^+$ &The dispersion parameter for removed cases\\
		\bottomrule
	\end{tabular}
	\label{sir_para_tab}
\end{table}

\clearpage
\newpage
\setcounter{section}{0}
\section{Supplement to Section 3: Posterior sampling {via} MCMC}\label{s3}
In this section, we discuss numerical procedures to obtain posterior samples for parameters of interest in the proposed model when the number of change points $M$ is pre-specified. Our approach is to develop a Metropolis-Hastings algorithm that draws samples from proposal distributions and accepts/rejects them based on the Hastings ratio. However, setting up a proposal for $\bm{\delta}$ is non-trivial as there is no known distribution that could sample a change-point indicator vector. Hence, we develop some moves in order to update the positions of elements of $\bm{\delta}$ without having to draw a sample from a known density. To update $\bm{\delta}$, we apply a combination of {local swap}, {global swap}, and {shift} moves. Performing the {local swap} step, we randomly select a change point at time $t$, say $\delta_t = 1$ and swap the entries at time $t$ and $(t\pm1)$, giving us the proposed indicator vector $\bm{\delta}^*$ and segmentation vector $\bm{z}^*$ such that $z_t^* = \sum_{t' = 1}^t \delta^*_{t'}$. Similarly, the {global swap} move results in randomly swapping a change point at time $t$ and a non-change point at time $t'$, such that $\delta_t = 0$ and $\delta_t' = 1$. This allows the algorithm to escape local maxima and explore other regions of interest. Finally, the {shift} move takes all the change points into consideration and shifts them by one-time point to the left or the right, with equal probability. The {local swap} and {shift} moves are performed with probability $0.4$ each, while the {global swap} move is performed with probability $0.2$. At each MCMC iteration, we accept the proposed indicator vector $\bm{\delta}^*$ with probability $\min(1, h)$, where $h$ is the Hastings ratio obtained as follows:
\begin{align}\label{1.1}
    h \quad = \quad \frac{f(\bm{\dot{C}}|\bm{\delta}^*,\bm{\Theta},\phi)}{f(\bm{\dot{C}}|\bm{\delta},\bm{\Theta},\phi)}\frac{\pi\left(\bm{\delta}^*\right)}{\pi\left(\bm{\delta}\right)}\frac{J\left({\bm{\delta}}\leftarrow{\bm{\delta}}^*\right)}{J\left({\bm{\delta}^*\leftarrow\bm{\delta}}\right)}.
\end{align}
Note that the last term of the proposal density ratio cancels out as all the moves result in equal probability in the numerator and denominator. 

Now we update each epidemiological parameter $\theta_m\in\bm{\Theta}_m = \{\lambda_m, K_m, p_m\}, m = 1,\cdots, M$ using appropriate proposal distributions. First, we draw $\ln \lambda^*_m \in \mathbb{R}$ from a normal distribution, i.e., $\ln \lambda^*_m \sim \text{N}(\ln \lambda_m, \sigma_{\lambda}^2)$ with the mean from the previous iteration given by $\ln \lambda_m$ and pre-determined step-size variance $\sigma_{\lambda}^2$. Updates in logarithmic scale allow efficient symmetric proposals for positively bounded parameters. The transformed parameter draws are exponentiated back to the original scale once the proposal is accepted. Then, we draw $\ln K^*_m$ from a truncated normal distribution, i.e., $\ln K^*_m \sim \text{TN}(\ln K_m, \sigma_K^2, \underset{t:z^*_t = m^*}{\max}\ln C_t, \ln\lceil \rho N \rceil)$, where the four terms within $\text{TN}(\cdot)$ represent the mean, variance, lower, and upper bounds, respectively. Last, $\ln p^*_m \in (-\infty, 0]$ is drawn from a truncated normal density such that $\ln p^*_m \sim \text{TN}(\ln p_m, \sigma_p^2, -\infty, 0)$.
At each MCMC iteration, we accept the proposed parameter $\theta^*_{m}$ with probability $\min(1, h)$, where $h$ is the Hastings ratio obtained as follows:
\begin{align}\label{1.2}
    h \quad = \quad \frac{f(\bm{\dot{C}}|\bm{\delta},\bm{\Theta}_{m},\phi)}{f(\bm{\dot{C}}|\bm{\delta},\bm{\Theta}_m,\phi)}\frac{\pi\left(\theta_{m}^*\right)}{\pi\left(\theta_{m}\right)}\frac{J\left({\theta_{m}}\leftarrow{\theta_{m}^*}\right)}{J\left({\theta_{m}^*\leftarrow\theta_{m}}\right)}.
\end{align}
Note that the proposal ratios are exactly equal to one for updating $\lambda_m$, while approximately equal to one for updating $K_m$ and $p_m$ as log transformations on parameters inject symmetry to the proposal distributions, resulting in efficient random-walk Metropolis samplers. 

Updating $\phi$ will follow a similar setup, replacing partial data likelihood with the complete data likelihood and proposing $\ln \phi^* \sim \text{TN}(\ln \phi, \sigma_\phi^2, 0, \ln 100)$ under reasonable constraints.

\section{Supplement to Section 4: Posterior sampling {via} RJMCMC}
The birth–death form of RJMCMC proposes a new parameter vector by first randomly choosing to increase the dimension of the parameter space by one (a birth), decrease the dimension by one (a death), or keep the dimension the same (a stay). In the case of a birth, a new component is added to the model according to a chosen distribution. Similarly, a component is "killed" through random selection. This extra step of selecting a move type and developing the proposal based on the selected move is accounted for in the acceptance probability of a proposal. Amid the trans-dimensional moves, we also perform a few fixed-dimensional moves (global swap, local swap, and stay moves) to explore the parameter space as efficiently as possible. The general Hastings ratio can be written as:
\begin{align}\label{move}
    h\quad =&\quad \frac{\pi(\bm{\delta}^*, \bm{\mathcal{T}}^*, M^*| \bm{\dot{C}}, \phi)}{\pi(\bm{\delta}, \bm{\mathcal{T}}, M| \bm{\dot{C}}, \phi)}\frac{J(\mathcal{M} \leftarrow \mathcal{M}^*)}{J(\mathcal{M}^* \leftarrow \mathcal{M})}\\
    \quad =&\quad \frac{\pi(\bm{\delta}^*, \bm{\mathcal{T}}^*, M^*| \bm{\dot{C}}, \phi)}{\pi(\bm{\delta}, \bm{\mathcal{T}}, M| \bm{\dot{C}}, \phi)}\frac{J(\bm{\mathcal{T}}|\bm{\mathcal{T}}^*, M^*)}{J(\bm{\mathcal{T}}^*|\bm{\mathcal{T}}, M)} \frac{J(\bm{\delta}|\bm{\delta}^*, M^*)}{J(\bm{\delta}^*|\bm{\delta}, M)}\frac{J(M|M^*)}{J(M^*|M)},
\end{align}
while the specific version for each move is:
\begin{align}
    h_\text{birth}\quad =& \quad \frac{\prod_{u=1}^{M + 1} \prod_{\{t:z^*_t=u\}} \frac{\Gamma(\dot{C}_t+\phi)}{\dot{C}_t!\Gamma(\phi)}\left(\frac{\phi}{g(C_{t-1},\bm{\Theta}^*_{u}) + \phi}\right)^{\phi}\left(\frac{g(C_{t-1},\bm{\Theta}^*_{u})}{g(C_{t-1},\bm{\Theta}^*_{u})+\phi}\right)^{\dot{C}_t}}{\prod_{v=1}^M \prod_{\{t:z_t=v\}} \frac{\Gamma(\dot{C}_t+\phi)}{\dot{C}_t!\Gamma(\phi)}\left(\frac{\phi}{g(C_{t-1},\bm{\Theta}_v) + \phi}\right)^{\phi}\left(\frac{g(C_{t-1},\bm{\Theta}_v)}{g(C_{t-1},\bm{\Theta}_v)+\phi}\right)^{\dot{C}_t}}\nonumber\\
    \quad \times& \quad \frac{\eta\omega_t^{\delta_t^*}(1 - \omega_t)^{1 - \delta_t^*}{\lambda^*_{m^*}}^{a_\lambda - 1}\exp{(b_\lambda \lambda^*_{m^*})}{p^*_{m^*}}^{a_p - 1}\exp{(b_p p^*_{m^*})}\left(\underset{\{t:z^*_{t} = m^*\}}{\max}C_t - \lceil \rho N \rceil\right)}{(M + 1)\omega_t^{\delta_t}(1 - \omega_t)^{1 - \delta_t}}\nonumber\\
    \quad \times& \quad \text{\tiny$\frac{T - M}{M}\left[\frac{\exp\left(-\frac{1}{2}\left\{ \frac{\left(\ln(K^*_{m^*}) - \ln(K_m)\right)^2}{\sigma_K^2} + \frac{\left(\ln(p^*_{m^*}) - \ln(p_m)\right)^2}{\sigma_p^2} + \frac{\left(\ln(\lambda^*_{m^*}) - \ln(\lambda_m)\right)^2}{\sigma_\lambda^2}\right\}\right)}{(2\pi)^{3/2}\sigma_K\sigma_p\sigma_\lambda\left[\Phi\left(\ln\lceil \rho N \rceil; \ln K_m, \sigma_K^2\right) - \Phi\left(\ln \underset{\{t:z_{t} = m\}}{\max}C_t; \ln K_m, \sigma_K^2\right)\right]\left[\Phi\left(0; \ln p_m, \sigma_p^2\right) - \Phi\left(-\infty; \ln p_m, \sigma_p^2\right)\right]}\right]^{-1}$} 
\end{align}
Here, $\bm{\Theta}^*_{m^*} = \{K^*_{m^*}, \lambda^*_{m^*}, p^*_{m^*}\}$ is the set of parameters proposed for the added new segment ($m^* = m + 1$). While, $\bm{\Theta}_{m} = \{K_{m}, \lambda_{m}, p_{m}\}$ is the old set of parameters for the original segment.
\begin{align}
    h_\text{death}\quad =&\quad \frac{\prod_{u=1}^{M - 1} \prod_{\{t:z^*_t=u\}} \frac{\Gamma(\dot{C}_t+\phi)}{\dot{C}_t!\Gamma(\phi)}\left(\frac{\phi}{g(C_{t-1},\bm{\Theta}^*_{u}) + \phi}\right)^{\phi}\left(\frac{g(C_{t-1},\bm{\Theta}^*_{u})}{g(C_{t-1},\bm{\Theta}^*_{u})+\phi}\right)^{\dot{C}_t}}{\prod_{v=1}^M \prod_{\{t:z_t=v\}} \frac{\Gamma(\dot{C}_t+\phi)}{\dot{C}_t!\Gamma(\phi)}\left(\frac{\phi}{g(C_{t-1},\bm{\Theta}_v) + \phi}\right)^{\phi}\left(\frac{g(C_{t-1},\bm{\Theta}_v)}{g(C_{t-1},\bm{\Theta}_v)+\phi}\right)^{\dot{C}_t}}\nonumber\\
    \quad \times& \quad \frac{M\omega_t^{\delta_t^*}(1 - \omega_t)^{1 - \delta_t^*}}{\eta\omega_t^{\delta_t}(1 - \omega_t)^{1 - \delta_t}\lambda_{m}^{a_\lambda - 1}\exp{(b_\lambda \lambda_{m})}p_{m}^{a_p - 1}\exp{(b_p p_{m})}\left(\underset{\{t:z_{t} = m\}}{\max}C_t - \lceil \rho N \rceil\right)}\nonumber\\
    \quad \times& \quad \text{\fontsize{4}{4}\selectfont$\frac{M - 1}{T - M + 1} \left[\frac{\exp\left(-\frac{1}{2}\left\{ \frac{\left(\ln(K_{m}) - \ln(K^*_{m^*})\right)^2}{\sigma_K^2} + \frac{\left(\ln(p_{m}) - \ln(p^*_{m^*})\right)^2}{\sigma_p^2} + \frac{\left(\ln(\lambda_{m}) - \ln(\lambda^*_{m^*})\right)^2}{\sigma_\lambda^2}\right\}\right)}{(2\pi)^{3/2}\sigma_K\sigma_p\sigma_\lambda\left[\Phi\left(\ln\lceil \rho N \rceil; \ln K^*_{m^*}, \sigma_K^2\right) - \Phi\left(\ln \underset{\{t:z^*_{t} = m^*\}}{\max}C_t; \ln K^*_{m^*}, \sigma_K^2\right)\right]\left[\Phi\left(0; \ln p^*_{m^*}, \sigma_p^2\right) - \Phi\left(-\infty; \ln p^*_{m^*}, \sigma_p^2\right)\right]}\right]$}
\end{align}
Here, $\bm{\Theta}^*_{m^*} = \{K^*_{m^*}, \lambda^*_{m^*}, p^*_{m^*}\}$ is the set of parameters of the new segment proposed after deletion of the $m^{th}$ segment such that $m^* = m - 1$. While, $\bm{\Theta}_{m} = \{K_{m}, \lambda_{m}, p_{m}\}$ is the old set of parameters of the deleted segment.

\noindent To derive $h_\text{global swap}$ and $h_\text{local swap}$, we successively update $\bm{\delta}$ and $\bm{\mathcal{T}}$ via Equations (\ref{1.1}) and (\ref{1.2}) respectively, for fixed $M$. Meanwhile, $h_\text{stay}$ updates $\bm{\mathcal{T}}$ via Equation (\ref{1.2}), for fixed $\bm{\delta}$ and $M$.

\section{Supplement to Section 5: Generative scheme based on the stochastic SIR model}\label{s5}

At any given time point $t$ in a closed population of size $N$, an individual belongs to one of the three compartments, namely, susceptible $(S_t)$, infectious $(I_t)$ or removed (dead + recovered, $R_t)$ such that $S_t + I_t + R_t = N $. Assuming $100\%$ population is susceptible at $t = 0$, 
an initial number of infectious individuals transmitted the disease at a disease transmission rate $\beta$. Subsequently, affected individuals may further transmit the disease and ultimately recover or die at a removal (death or recovery) rate $\gamma$. Another metric of great importance is the basic reproduction number $\mathcal{R}_0 = \beta/\gamma$ that measures the expected number of individuals likely to be infected by a single disease carrier in a closed population. The cumulative case number at a given time point is expressed as an aggregate of infectious and removed individuals. Hence, we have $C_t = N - S_t$. To set up the simulation study, we considered $T = 120$ time points divided across $M = 4$ sub-segments and a fixed population $N = 1,000,000$. The equally spaced change points were assumed at $t = 31, 61, 91$. For the parameter settings, segment-specific basic reproduction numbers and removal rate were specified to be $\bm{\mathcal{R}}_0 = (3.0, 2.0, 1.1, 0.5)^\top$ and $\gamma = 0.03$, respectively. Segment-specific transmission rates were simply obtained by $\bm{\beta} = \bm{\mathcal{R}}_0/\gamma$. Such a scenario was characterized by a gradually decreasing disease transmission rate due to successful control over the epidemic. An accurate change-point detection model must distinguish between the subtleties of an actual change-point and variations due to chance or pure error. To control the measurement error, we considered scenarios of high and low dispersion attributed by $\phi_S = \phi_R = 10$ and $\phi_S = \phi_R = 100$, respectively. The stochastic SIR model samples $S_t$ and $R_t$ from a NB distribution, while $I_t$, $C_t$ and $\dot{C}_t$ were obtained sequentially as follows: 
\[
\begin{cases}
 S_t = S_{t - 1} - \text{NB}\left(\beta_m N^{-1}S_{t - 1} I_{t - 1}, \phi_S\right)\\
 R_t = R_{t - 1} + \text{NB}\left(\gamma I_{t - 1}, \phi_R\right)\\
 I_t = N - S_t - R_t\\
 C_t = N - S_t\\
 \dot{C}_t = -(S_t - S_{t - 1}) = C_t - C_{t - 1}
\end{cases},
\]
for all $t = 1,\cdots, T$, $m = 1,\cdots, M$, where we assumed $I_0 = 100$, $R_0 = 0$ and hence, $C_0 = 100$. We repeated the above steps to generate $50$ independent simulated datasets for each setting.

The key notations of the stochastics SIR model are listed in Table S2.

\section{Supplement to Section 5: Evaluation metrics}\label{s6}
 Let $\text{a}=\sum_{t > t'}\delta(z_t = z_{t'})\delta(\hat{z}_{t} = \hat{z}_{t'})$; $\text{b}=\sum_{t > t'}\delta(z_t = z_{t'}) \delta(\hat{z}_{t} \neq \hat{z}_{t'})$; $\text{c}=\sum_{t > t'}\delta(z_t \neq z_{t'}) \delta(\hat{z}_{t} = \hat{z}_{t'})$; and $\text{d}=\sum_{t > t'}\delta(z_t \neq z_{t'}) \delta(\hat{z}_{t} \neq \hat{z}_{t'})$ be the number of pairs of time points that are: a) in the same segment in both of the true and estimated partitions; b) in different segments in the true partition but in the same segment of the estimated one; c) in the same segment of the true partition but in different segments in the estimated one; and d) in different segments in both of the true and estimated partitions. Further let $n_m = \sum_{t = 1}^T I(z_t = m)$, $\hat{n}_{m'} = \sum_{t = 1}^T I(\hat{z}_t = m')$ be the lengths of the true sub-epidemic $m$ and the estimated sub-epidemic $m'$ respectively and $n_{mm'} = \sum_{t = 1}^TI(z_t = m, \hat{z}_t = m')$ be the number of correct classifications between the true and estimated segments
 
\subsection{Adjusted Rand index}
Then, the ARI can be computed as:
\begin{align*}
\text{ARI}(\bm{z}, \hat{\bm{z}})\quad=\quad\frac{\binom{T}{2}(\text{a} + \text{d}) - [(\text{a} + \text{b})(\text{a} + \text{c}) +(\text{c} + \text{d})(\text{b} + \text{d})]}{\binom{T}{2}^2 - 
	[(\text{a} + \text{b})(\text{a} + \text{c}) +(\text{c} + \text{d})(\text{b} + \text{d})]}.
\end{align*}
The ARI usually yields values between $0$ and $1$, although it can yield negative values \citep{hubert1985comparing}. The larger the index, the more similarities between $\bm{z}$ and $\hat{\bm{z}}$, and thus the more accurately the method detects the actual times at which change points occurred.

\subsection{Mutual information}
 $\text{MI}$ \citep{steuer2002mutual} measures the amount of information one clustering shares with the other. MI can be computed as:
\begin{align}
    \text{MI}(\bm{z}, \hat{\bm{z}}) &= \frac{1}{T}\sum_{m = 1}^M\sum_{m' = 1}^M n_{mm'} \log\left(\frac{n_{mm'}T}{n_m\hat{n}_{m'}}\right).
\end{align}
It yields non-negative values. The larger the MI, the more accurate the partition result.

\subsection{Normalized variation of information}
\cite{meilua2007comparing} proposed a metric namely {Variation of Information} $(\text{VI})$ that takes into account the amount of information lost in choosing $\hat{\bm{z}}$ over $\bm{z}$, along with the information required by the former to match with the latter. Apart from the useful axiomatic properties of this metric, \cite{meilua2007comparing} stated the upper bound of the metric given by $\text{VI}(\bm{z}, \hat{\bm{z}}) \leq \log(T)$. Hence, we consider the {Normalized Variation of Information} (NVI),
\begin{align}
    \text{NVI}(\bm{z}, \hat{\bm{z}}) &= -\frac{1}{T\log(T)}\left[\sum_{m = 1}^Mn_m\log\left(\frac{n_m}{T}\right) + \sum_{m' = 1}^M\hat{n}_{m'}\log\left(\frac{\hat{n}_{m'}}{T}\right)\right.\\
    & \quad + \left. 2 \sum_{m = 1}^M\sum_{m' = 1}^M n_{mm'} \log\left(\frac{n_{mm'}T}{n_m\hat{n}_{m'}}\right)\right],
\end{align}
where NVI $ \in [0,1]$ such that NVI $ = 0$ refers to perfect similarity as there is no information lost by selecting $\hat{\bm{z}}$ over $\bm{z}$. Likewise, there is no lost information in $\hat{\bm{z}}$. On the other hand, having NVI $ = 1$ is a consequence of perfect dissimilarity (i.e., the entire information is lost in estimating $\bm{z}$ and no information on the true clustering is gained).

\subsection{F-measure}

The F-measure for clustering \citep{fung2003hierarchical} is based on computing the weighted harmonic mean of {precision} and {recall}. Where, perfect {precision} would result in a cluster with only the correct labels but not necessarily all the correct ones. On the contrary, perfect {recall} would be a consequence of considering all possible true labels but also including the incorrect ones. Let the true segmentation and estimated segmentation vectors having $M$ sub-segments each, be denoted by $\bm{z}$ and $\hat{\bm{z}}$. Then the F-measure for the two vectors is given by:
\begin{align}
\text{F}(\bm{z}, \hat{\bm{z}}) 
&= \frac{2}{T}\sum_{m = 1}^M n_m\; \underset{m' = \{1,\cdots,M\}}{\max} \left\{\frac{n_{mm'}}{n_m + \hat{n}_{m'}}\right\},
\end{align}
 Here, F $ \in [0,1]$ such that larger values of F$(\bm{z}, \hat{\bm{z}})$ are a consequence of better similarity between the actual and estimated segmentation.

\bibliographystyle{agsm}
{\footnotesize
	\bibliography{aoas_supp.bib}}